\documentclass[useAMS,usenatbib,usegraphicx]{aa}
\usepackage{txfonts}
\usepackage{epsfig}
\usepackage{subfigure}
\usepackage{multirow}

\usepackage{times}
\usepackage{natbib}
\usepackage{longtable,lscape}
\bibpunct{(}{)}{;}{a}{}{,}

\usepackage[bookmarks=false]{hyperref}

\def\ltsima{$\; \buildrel < \over \sim \;$}
\def\simlt{\lower.5ex\hbox{\ltsima}}
\def\gtsima{$\; \buildrel > \over \sim \;$}
\def\simgt{\lower.5ex\hbox{\gtsima}}
\def\gsimeq
{\hbox{\raise0.5ex\hbox{$>\lower1.06ex\hbox{$\kern-1.07em{\sim}$}$}}}
\def\lsimeq
{\hbox{\raise0.5ex\hbox{$<\lower1.06ex\hbox{$\kern-1.07em{\sim}$}$}}}

\def\xmm{{\it XMM-Newton }}

\def\xmm{{\it XMM-Newton}}
\def\chandra{{\it Chandra}}
\def\suzaku{{\it Suzaku}}
\def\xspec{{\it Xspec}}

\def\apj{ApJ}
\def\mnras{MNRAS}
\def\aap{A\&A}
\def\apjl{ApJ}
\def\apjs{ApJS}
\def\araa{ARA\&A}
\def\pasj{PASJ}

\def\xis{XIS}
\def\xis1{XIS1}
\def\xis2{XIS2}
\def\xis3{XIS3}

\def\Fevc{Fe~{\sc xxv}}
\def\Fevs{Fe~{\sc xxvi}}
\def\FeKa{Fe~K$\alpha$}

\title{Multiwavelength campaign on Mrk 509: \\
 XI. Reverberation of the Fe K$\alpha$ line}

\author{G. Ponti\inst{1}
  \and M. Cappi\inst{2}
  \and E. Costantini\inst{3}
  \and S. Bianchi\inst{4}
  \and J.S. Kaastra\inst{3,5}
  \and B. De Marco\inst{2,6}
  \and R. P. Fender\inst{1}
  \and P.-O. Petrucci\inst{7}
  \and G.A. Kriss\inst{8,9}
  \and K.C. Steenbrugge\inst{10,11}
  \and N. Arav\inst{12}
  \and E. Behar\inst{13}
  \and G. Branduardi-Raymont\inst{14}
  \and M. Dadina\inst{2}
  \and J. Ebrero\inst{3}
  \and P. Lubi\'nski\inst{15}
  \and M. Mehdipour\inst{14}
  \and S. Paltani\inst{16}
  \and C. Pinto\inst{3}
  \and F. Tombesi\inst{17,18}
  }

\institute{School of Physics and Astronomy, University of Southampton, Highfield, Southampton SO17 1BJ
	   \and 
	   INAF-IASF Bologna, Via Gobetti 101, 40129 Bologna, Italy
	   \and	   
	   SRON Netherlands Institute for Space Research, Sorbonnelaan 2, 3584 CA Utrecht, the Netherlands 
	   \and 
	   Dipartimento di Fisica, Universit\`a degli Studi Roma Tre, via della Vasca Navale 84, 00146 Roma, Italy 
	   \and	   
	   Sterrenkundig Instituut, Universiteit Utrecht, P.O. Box 80000, 3508 TA Utrecht, the Netherlands
	   \and 
	   Centro de Astrobiolog\'ia (CSIC-INTA), Dep. de Astrof\'isica; LAEFF, PO Box 78, E-28691, Villanueva de la Ca\~nada, Madrid, Spain
	   \and
	   UJF-Grenoble 1 / CNRS-INSU, Institut de Plan\'etologie et d'Astrophysique de Grenoble (IPAG) UMR 5274, Grenoble, F-38041, France
           \and 
	   Space Telescope Science Institute, 3700 San Martin Drive, Baltimore, MD 21218, USA
	   \and 
	   Department of Physics and Astronomy, The Johns Hopkins University, Baltimore, MD 21218, USA
	   \and 
           Instituto de Astronom\'ia, Universidad Cat\'olica del Norte, Avenida Angamos 0610, Casilla 1280, Antofagasta, Chile
	   \and 
	   Department of Physics, University of Oxford, Keble Road, Oxford OX1 3RH, UK
	   \and 
	   Department of Physics, Virginia Tech, Blacksburg, VA 24061, USA
	   \and 
	   Department of Physics, Technion-Israel Institute of Technology, Haifa 32000, Israel 	
          \and
	   Mullard Space Science Laboratory, University College London, Holmbury St. Mary, Dorking, Surrey, RH5 6NT, UK
	   \and 
	   Centrum Astronomiczne im. M. Kopernika, Rabia\'nska 8, PL-87-100 Toru\'n, Poland
	   \and
	   ISDC Data Centre for Astrophysics, Astronomical Observatory of the University of Geneva, 16, ch. d'Ecogia, 1290 Versoix, Switzerland 
	   \and 
           Department of Astronomy and CRESST, University of Maryland, College Park, MD 20742, USA
           \and 
           X-ray Astrophysics Laboratory, NASA/Goddard Space Flight Center, Greenbelt, MD, 20771, USA
} 

\offprints{Gabriele Ponti\\ \email{ponti@iasfbo.inaf.it}}

\authorrunning{G. Ponti et al.}

\usepackage{times}

\begin{document}

\abstract{We report on a detailed study of the Fe K emission/absorption complex in 
the nearby, bright Seyfert 1 galaxy Mrk~509. The study is part of an extensive 
\xmm\ monitoring consisting of 10 pointings ($\sim60$ ks each) about once every 
four days, and includes also a reanalysis of previous \xmm\ and \chandra\ 
observations.}
{We aim at understanding the origin and location of the Fe K emission 
and absorption regions.}
{We combine the results of time-resolved spectral analysis on both short and 
long time-scales including model independent rms spectra.}
{Mrk~509 shows a clear (EW$=58\pm4$ eV) neutral \FeKa\ emission line that can be 
decomposed into a narrow ($\sigma=0.027$ keV) component (found in the \chandra\ 
HETG data) plus a resolved ($\sigma=0.22$ keV) component. 
We find the first successful measurement of a linear correlation between the intensity 
of the resolved line component and the 3-10 keV flux variations on time-scales 
of years down to a few days. The Fe K$\alpha$ reverberates the hard X-ray 
continuum without any measurable lag, suggesting that the region 
producing the resolved \FeKa\ component is located within a few light 
days-week (r $\lsimeq~10^3$ r$_{\rm g}$) from the Black Hole (BH). 
The lack of a redshifted wing in the line poses a lower limit of $\geq$40 r$_{\rm g}$ 
for its distance from the BH. The Fe K$\alpha$ could thus be emitted from the 
inner regions of the BLR, i.e. 
within the $\sim$80 light days indicated by the H$\beta$ line measurements. 
In addition to these two neutral \FeKa\ components, we confirm the detection 
of weak (EW$\sim8-20$ eV) ionised Fe K emission. This ionised line 
can be modeled with either a blend of two narrow \Fevc\ and \Fevs\ emission lines 
(possibly produced by scattering from distant material) or with a single 
relativistic line produced, in an ionised disc, down to a few r$_{\rm g}$ from the BH. 
In the latter interpretation, the presence of an ionised standard $\alpha$-disc,
down to a few r$_{\rm g}$, is consistent with the source high Eddington ratio.
Finally, we observe a weakening/disappearing of the medium and high velocity 
high ionisation Fe K wind features found in previous \xmm\ observations.}
{This campaign has made possible the first reverberation measurement of the 
resolved component of the Fe K$\alpha$ line, from which we can infer a location 
for the bulk of its emission at a distance of r$\sim40-1000$ r$_{\rm g}$ 
from the BH.}

\maketitle

\section{Introduction}

X-ray observations of AGN have shown the almost 
ubiquitous presence of the \FeKa\ line at 6.4 keV (Yaqoob et al. 2004; 
Nandra et al. 1997; 2007; Bianchi et al. 2007; de la Calle et al. 2010). 
Unlike the optical-UV lines that are emitted 
by distant material only, the \FeKa\ line traces reflection not only from distant 
material (such as the inner wall of the molecular torus, the 
broad line region and/or the outer disc) but also from regions as close  
as a few r$_{\rm g}$ (where r$_{\rm g}$=GM/c$^2$) from the BH (Fabian et al. 2000).

The powerful reverberation mapping technique, routinely exploited 
on optical-UV lines (Clavel et al. 1991; Peterson 1993; Kaspi et al. 2000; 
Peterson et al. 2004), can also be applied to X-ray lines such as the \FeKa\ line. 
This kind of analysis has a tremendous potential, allowing us to map 
the geometry of matter surrounding the BH, starting from distances of a few 
gravitational radii up to light years. However, each \FeKa\ component is expected to 
respond on a different characteristic time (years-decades for the torus, 
several days-months for the BLR-outer disc, and tens of seconds to a few hours 
for the inner accretion disc) and current X-ray instruments cannot easily disentangle
the different components. Indeed, reverberation mapping of all \FeKa\ emission 
components represents an enormous observational challenge,
and specially tailored monitoring campaigns (to sample the proper time scales) 
have to be designed.

Since the detection of the first clear example of a broad and skewed Fe line profile 
in the spectrum of an AGN (indicating that most of the line emission is produced 
within few tens of r$_{\rm g}$; e.g. MCG-6-30-15, Tanaka et al. 1995) 
the quest to understand how the broad \FeKa\ line varies with the continuum is ongoing. 
Indeed, close to the BH the simple one-to-one correlation between continuum and 
reflection line is distorted by General and Special relativistic effects. 
Several papers present extensive theoretical computations to describe the inner
disc reverberation to the continuum taking into account all relativistic 
effects (Reynolds et al. 1999; Fabian et al. 2000; Reynolds \& Nowak 2003). 

Several techniques have been employed to measure the variability-reverberation 
of the relativistic \FeKa\ line. However, for the best cases such as MCG-6-30-15, 
the relativistic Fe line showed a complex behaviour, having a variable intensity 
at low fluxes (Ponti et al. 2004; Reynolds et al. 2004) while showing a constant 
intensity at higher fluxes (Vaughan et al. 2003; 2004; see also the case of 
NG4051: Ponti et al. 2006). This puzzling and unexpected behaviour has 
been interpreted as due to strong light bending effects by some authors 
(Miniutti et al. 2003; 2004) or, alternatively, as the evidence that the broad 
wing of the \FeKa\ line is produced by strong and complex absorption 
effects (Miller et al. 2008). 

Thanks to the application of \FeKa\ excess emission maps (Iwasawa et al. 2005; 
Dovciak et al. 2004; De Marco et al. 2009), it has been possible to track 
weaker coherent patterns of \FeKa\ variations. In a few sources \FeKa\ 
variations are consistent with being produced by orbiting spots at a few r$_g$ from 
the BH (Iwasawa et al. 2004; Turner et al. 2006; Petrucci et al. 2007; 
Tombesi et al. 2007). Future larger area telescopes are needed to finally 
assess if these features are present only sporadically during peculiar periods 
or if, instead, although weak, are always present and can be used to map 
the inner disc (see e.g. Vaughan et al. 2008; De Marco et al. 2009). 

A leap forward in X-ray reverberation studies occurred thanks to the 
application of pure timing techniques to the long \xmm\ observation of 
1H0707-495 that allowed the discovery of a "reverberation lag" between 
the direct X-ray continuum and the soft excess, probably dominated by FeL  
line emission (Fabian et al. 2009; Zoghbi et al. 2010).
Soon after similar delays were seen in a few other objects (Ponti et al. 2010; 
De Marco 2011; Emmanouloulos et al. 2011; Zoghbi \& Fabian 2011; 
Turner et al. 2011). 
Recently, De Marco et al. (2012) showed that these lags are ubiquitous 
in AGN, that they scale with M$_{\rm BH}$ and have amplitudes of the order of 
the light crossing time of a few r$_g$, thus suggesting a reverberation 
origin of the delay (but see also Miller et al. 2010). 
Another fundamental step forward will be to combine these timing 
techniques to detect reverberation lags in the Fe K band (see Zoghbi 
et al. 2012). 

Reverberation from distant material has the advantage that the intensity of 
the \FeKa\ line and the continuum are expected to follow a simple one-to-one 
correlation, however, the expected delays between the reflection component 
and the direct emission are usually too large for a typical X-ray exposure. 
In fact, reflection from the inner walls of a molecular torus is expected to be delayed 
by a few years up to several decades and thus requires a very long monitoring campaign. 
Reflection from the BLR and/or outer disc is more accessible, 
the delay between continuum and reflection is expected to be between 
a few days up to few months. Thus a properly tailored monitoring campaign on 
a bright AGN with \xmm, \chandra\ or \suzaku\ could achieve this goal.
Several attempts have been made (Markowitz et al. 2003; Yaqoob et al. 
2005; Liu et al. 2010). However, the 15-20 \% or larger error on 
the flux of the \FeKa\ line and the low-sampling frequency of the X-ray 
observations have made the application of reverberation of the \FeKa\ line 
on weeks-months timescales basically impossible, until now. 

Mrk~509 (z=0.034397) is one of the brightest Seyfert 1 galaxies of the 
(2--100 keV) X-ray sky (Malizia et al. 1999; Revnivtsev et al. 2004; 
Sazonov et al. 2007), 
thus it has been observed by all major X-ray/Gamma-ray satellites.
The \chandra\ HETG spectrum shows a narrow 
component of the Fe K line with an Equivalent Width (EW) of 50 eV (Yaqoob 
et al. 2004). \xmm\ and \suzaku\ data provide evidence for a second broader 
($\sigma=0.12$ keV) neutral Fe K line (Ponti et al. 2009) 
as well as a weak ionized emission feature between 6.7--6.9 keV
(Pounds et al. 2001; Page et al. 2003; Ponti et al. 2009). The ionised 
emission can be fit either using a relativistically broadened ionised line 
or an outflowing photo-ionised gas component. 

Imprinted on the Fe K band emission of Mrk 509 are the fingerprints 
of two kinds of ionised absorption components, one marginally 
consistent with a medium velocity outflow (v$\sim14000$ km s$^{-1}$; 
Ponti et al. 2009) and the others out(in)flowing with relativistic velocities 
(Cappi et al. 2009; Dadina et al. 2005; Tombesi et al. 2010).

Here, we present the spectral and variability analysis of the Fe K complex 
energy band of Mrk~509 using the set of 10 \xmm\ observations (60 ks each), 
about one every fours days and spanning more than 1 month, which we 
obtained in 2009 (see the 3-10 keV light curve in Fig. \ref{lc}). We also  
re-analyse the previous 5 \xmm\ observations. 
Thanks to this extensive monitoring campaign we can measure correlated 
variations between the Fe K line intensity and X-ray continuum flux, 
allowing us, for the first time, to perform a reverberation mapping study 
on this X-ray emission line. In addition we can study the presence of highly 
ionised matter from the innermost regions around the BH. 

The paper is organised as follows. 
\S 2 is devoted to the description of the observations and 
data reduction. In \S 3 a first parametrisation (with a single Gaussian 
profile for the Fe K$\alpha$ line) of the total summed 
spectrum of the 2009 campaign is presented. 
Section 4 is dedicated to the detailed study of the \FeKa\ emission. 
We first present the study of the \FeKa\ line variability, assuming a single Gaussian 
profile (\S 4.1) we then use the \chandra\ HETG data (\S 4.2) to decompose the 
Fe K$\alpha$ line in two Gaussian (narrow and resolved) components. 
\S 4.3 presents the correlation between \FeKa\ intensity and the 3-10 keV 
continuum (once the Fe K$\alpha$
line is fitted with 2 Gaussian lines) which is confirmed,
in a model independent way, by the rms spectrum (\S 4.4). 
In \S 4.5 we discuss the possible origin of the \FeKa\ line.
\S 5 presents the study and the discussion of the origin of the 
ionised Fe K emission/absorption. Conclusions are in \S 6.

\section{Observations and data reduction}

\begin{figure}
\includegraphics[width=0.49\textwidth,height=0.35\textwidth,angle=0]{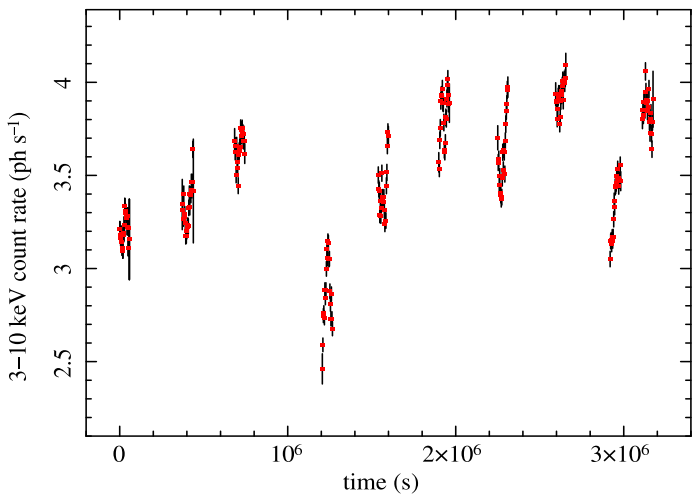}
 \caption{3-10 keV EPIC-pn light curve (3 ks time bins) of the 
 10 \xmm\ observations of the 2009 monitoring campaign.}
\label{lc}
\end{figure}

Mrk~509 was observed for a total of 15 times by \xmm: on 2000--10--25, 
2001--04--20, 2005--10--16, 2005--10--20, 2006--04--25, and 10 
times in 2009 (see Fig. \ref{lc} starting from 2009-10-15 and ending on 2009-11-20).
Ponti et al. (2009) and Kaastra et al. (2011) provide a full description of the first 5 
and the last 10 \xmm\ observations, respectively. 

We initially reduced the EPIC data (as in Mehdipour et al. 2011), starting from 
the ODF files, using the standard SAS v9.0 software. However, we noted that 
the rest-frame best fit energy of the \FeKa\ line in the EPIC-pn spectrum 
(E$_{\rm Fe K\alpha}=6.35\pm0.01$ keV) was not consistent with the best 
fit energy in the summed spectrum of the EPIC-MOS data (E$_{\rm Fe 
K\alpha}=6.41\pm0.01$ keV). 
This discrepancy ($\sim50$ eV) was found to be systematic, and 
was present in all 10 observations.
Being significantly larger than the reported systematic uncertainty on 
the calibration of the absolute energy scale of 10 eV (CAL-TN-0018), 
this result triggered an in-depth study of the pn and MOS energy scales 
by the \xmm\  EPIC calibration team.
After excluding that this effect is related to X-ray loading, 
a stronger than expected long-term degradation/evolution of the Charge 
Transfer Inefficiency (CTI) was found.  
The pn long-term CTI was thus re-calibrated, and its corrected value 
implemented in the SAS version 10.0.0 (see CCF release note
XMM-CCF-REL-271
\footnote{See http://xmm2.esac.esa.int/external/xmm\_sw\_cal/calib/rel\_notes/ index.shtml}).

The EPIC data were thus reduced again using the SAS version 10.0.0. 
During the XMM-Newton monitoring, both the EPIC-pn and the EPIC-MOS cameras 
were operating in the Small Window mode with the Thin filter applied. 
The \#XMMEA\_EP and \#XMMEA\_EM, for the pn and MOS cameras, respectively,
are used to filter the events lists and to create Good Time Intervals (GTI).
The FLAG==0 is then used for selection of events for making the spectra. 
The data were screened for increased flux of background particles. 
The contribution due to soft protons 
flares was negligible during the whole 2009 monitoring. The final cleaned EPIC-pn 
exposures for each XMM-Newton observation were about 60 ks, i.e. 
roughly 40 ks, after accounting for the proper dead-time of the pn when operating 
in small window mode (see Table 1 of Mehdipour et al. 2011, for a list of the 
exposure times).

The pn and MOS spectra were extracted from a circular region 
of $45''$ and $20''$ radius centred on the source, respectively. 
The background was taken locally from identical circular regions 
located on the same CCD of the source for the EPIC-pn but on another 
CCD for the EPIC-MOS.
The EPIC data showed no evidence of significant pile-up, thus single and  
double events were selected for both the pn (PATTERN$<=$4) and 
the MOS (PATTERN$<=$12) camera. Response matrices were generated  
for each source spectrum using the SAS tasks $arfgen$ and $rmfgen$.
The sum of the spectra has been performed with the {\sc 
mathpha}, {\sc addrmf} and {\sc addarf} tools within the 
{\sc Heasoft} package (version 6.10).

Mrk509 was observed by the Chandra Advanced CCD Imaging 
Spectrometer (ACIS: Garmire et al. 2003) with the High-Energy Transmission 
Grating Spectrometer (HETGS: Canizares et al. 2005) in the focal plane,
on 2001, April 13th (obsid 2087). 
Data were reduced with the Chandra Interactive Analysis of Observations
(CIAO: Fruscione et al. 2006) 4.2 and the Chandra Calibration DataBase 
(CALDB) 4.3.1 software, adopting standard procedures.

All spectral fits were performed using the Xspec software (version 12.3.0) 
and include the neutral Galactic absorption (4.44$\times$10$^{20}$~cm$^{-2}$; 
Murphy et al. 1996), the energies are in the rest frame if not specified otherwise, 
however the energies in the plots are in the observed frame and the errors 
are reported at the 90 per cent confidence level for one interesting 
parameter (Avni 1976) in all the tables, while they are 1 $\sigma$ 
errors in the Figures. 
Mrk 509 has a cosmological redshift of 0.034397 (Huchra 
et al. 1993) corresponding to a luminosity distance of 145 Mpc (taking 
$<H_0>=73$ km s$^{-1}$ Mpc$^{-1}$, $\Omega_{\Lambda}=0.73$ and 
$\Omega_{\rm m}=0.27$). 

\section{The mean spectrum}
\label{mean}

\begin{figure}
\hskip -0.8cm
\includegraphics[width=0.39\textwidth,height=0.61\textwidth,angle=-90]{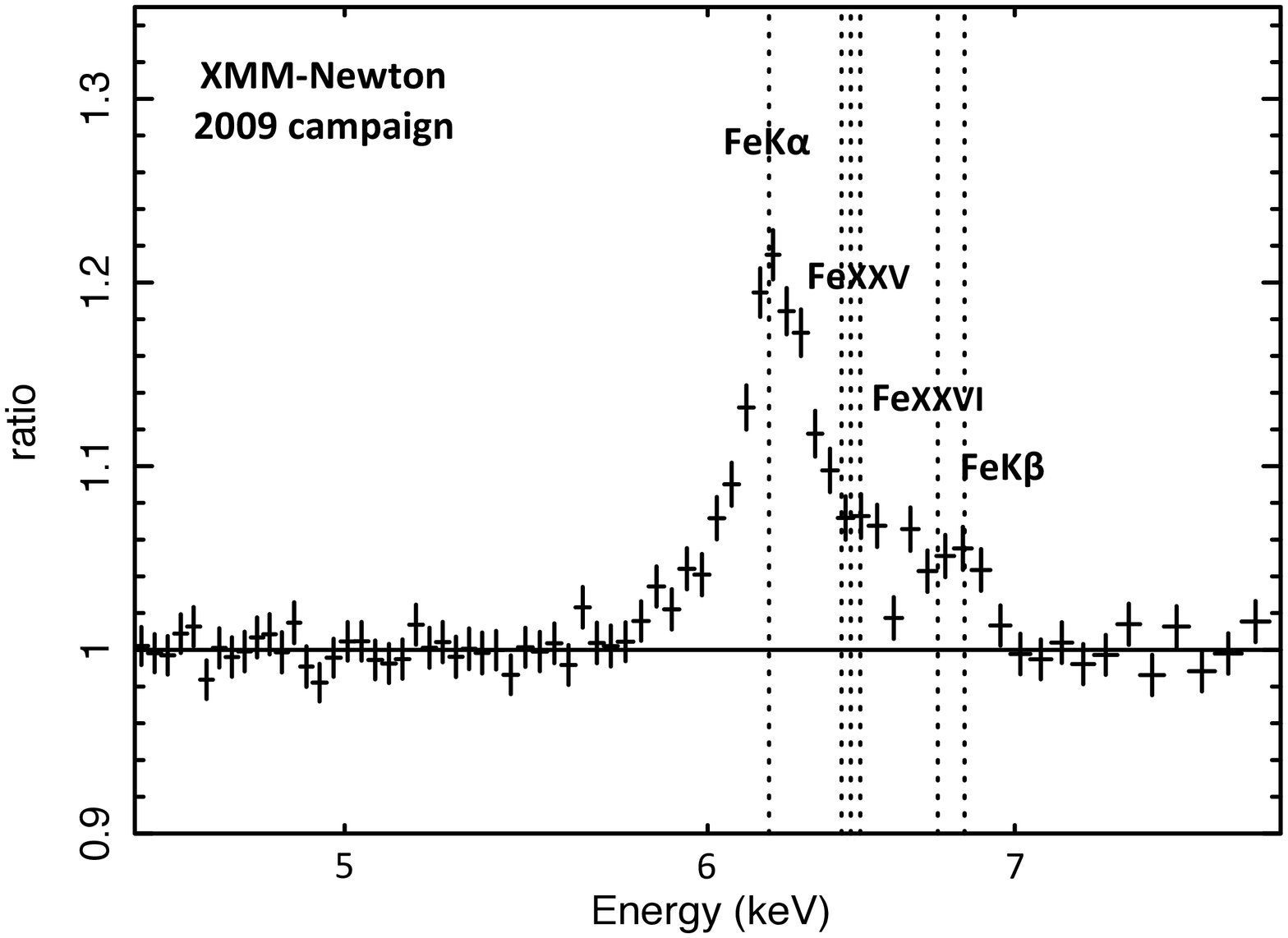}
\includegraphics[width=0.32\textwidth,height=0.46\textwidth,angle=-90]{rigaXmmSuzaku.ps} 
 \caption{{\it (Upper panel)} 
 Summed EPIC-pn spectra of the 10 observations performed during 
 the 2009 \xmm\ monitoring campaign. The data are fit, 
 in the 4-5 and 7.5-10 keV bands, with a simple power law, absorbed 
 by Galactic material, and the ratio of the data to the best-fit model is shown. 
 The dotted lines indicate the rest frame positions of the Fe K$\alpha$, 
 Fe K$\beta$, \Fevc\ (resonance, intercombination and forbidden) 
 and \Fevs\ lines. The x-axis reports the observed-frame energy. 
 {\it (Lower panel)} In red the 2006 XIS0+XIS3 \suzaku\ summed mean 
 spectra are shown.
 In black the summed spectra of the 5 \xmm\ observations performed between 2000 
 and 2006 are shown. This Figure is taken from Ponti et al. (2009). The arrows mark 
 possible absorption features in the spectrum.}
\label{FeKmean}
\end{figure}
The upper panel of Figure \ref{FeKmean} shows the data to best fit model 
ratio plot of the summed spectra of the 10 EPIC-pn observations performed 
during the 2009 \xmm\ monitoring campaign, fitted in the 3.5-5 and 7.5-10 keV band with 
a simple power law, absorbed by Galactic material (interstellar neutral gas; 
{\sc phabs} model in {\sc Xspec}). 
For comparison, the black data points in the lower panel of 
Fig. \ref{FeKmean} show the same plot for the summed EPIC-pn spectrum 
of the previous 5 \xmm\ observations taken between 2000 and 2006, 
while the red data show the summed XIS0+XIS3 spectra of 
the 4 \suzaku\ observations performed between April and November 
2006 (see Ponti et al. 2009 for more details).

Thanks to a longer integrated exposure and a slightly higher flux, the source spectrum 
in the FeK band has significantly better statistics during the 2009 campaign 
than the sum of all the previous observations (see Fig. \ref{FeKmean}). 
Hence we can better constrain the FeK complex and study its variability 
not only on the time-scales of days and weeks over which the 
monitoring has been performed, but also on time-scales of years, 
using previous observations. 
\begin{figure*}
\includegraphics[width=0.33\textwidth,height=0.25\textwidth]{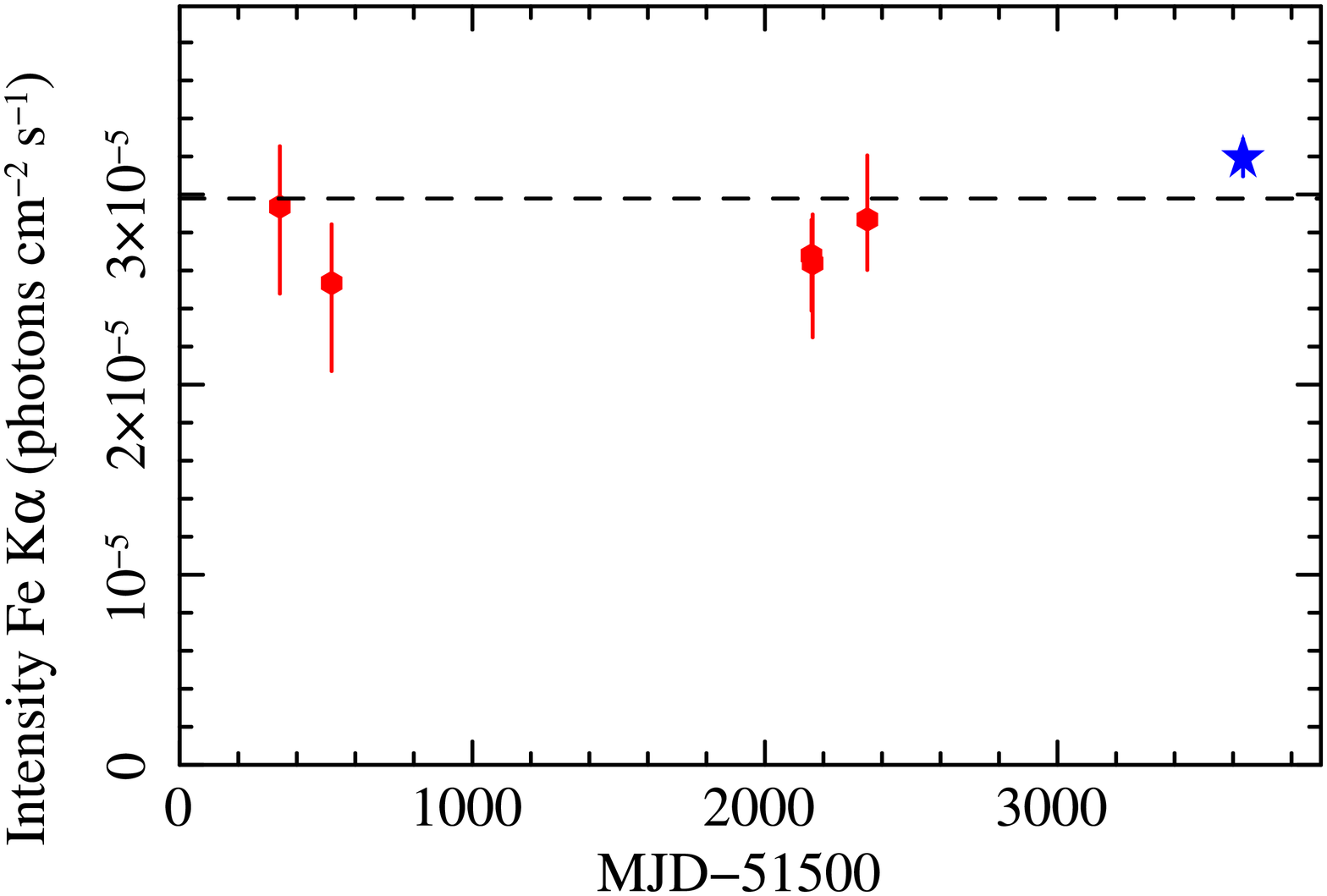}
\includegraphics[width=0.33\textwidth,height=0.25\textwidth]{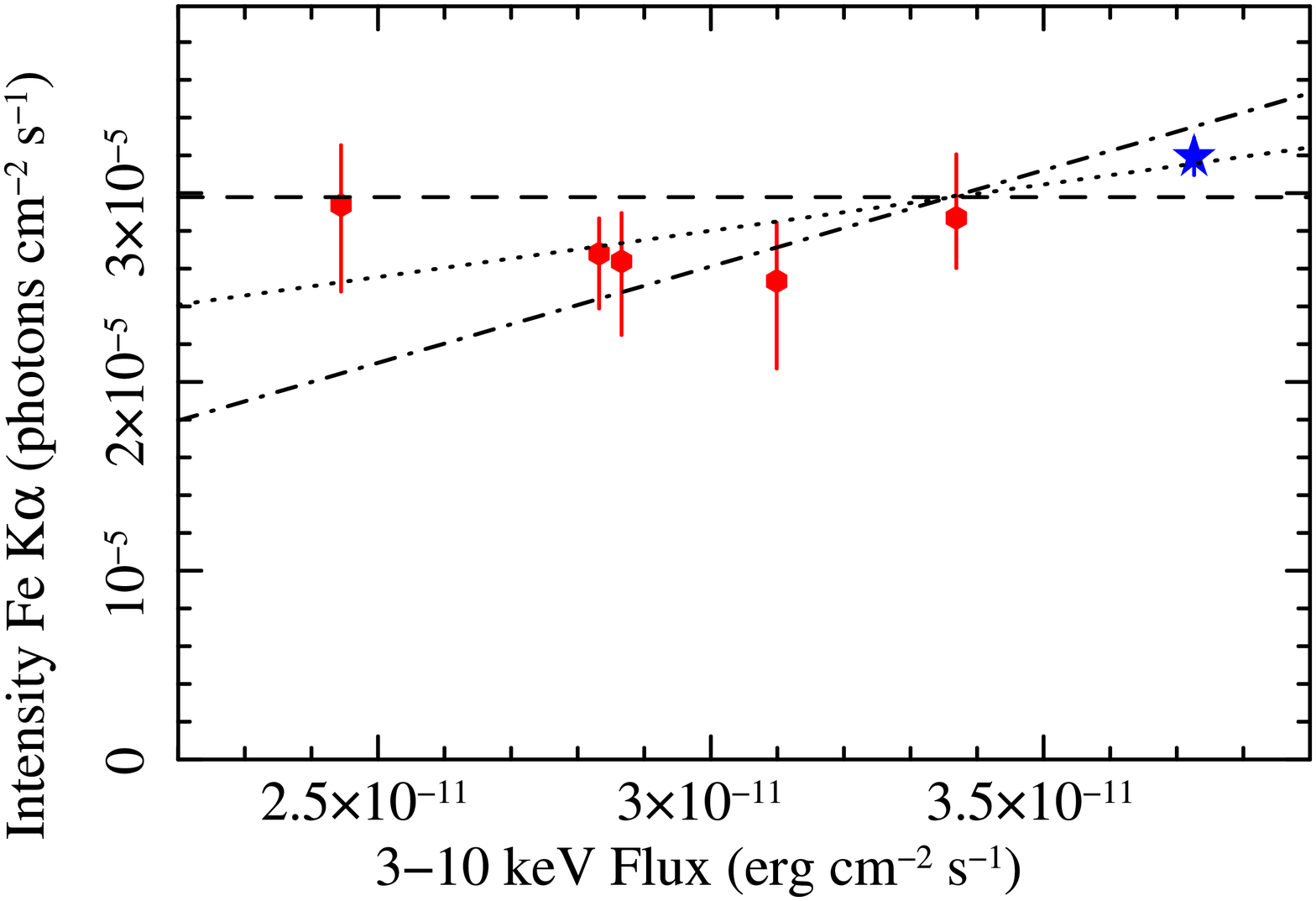}
\includegraphics[width=0.33\textwidth,height=0.25\textwidth]{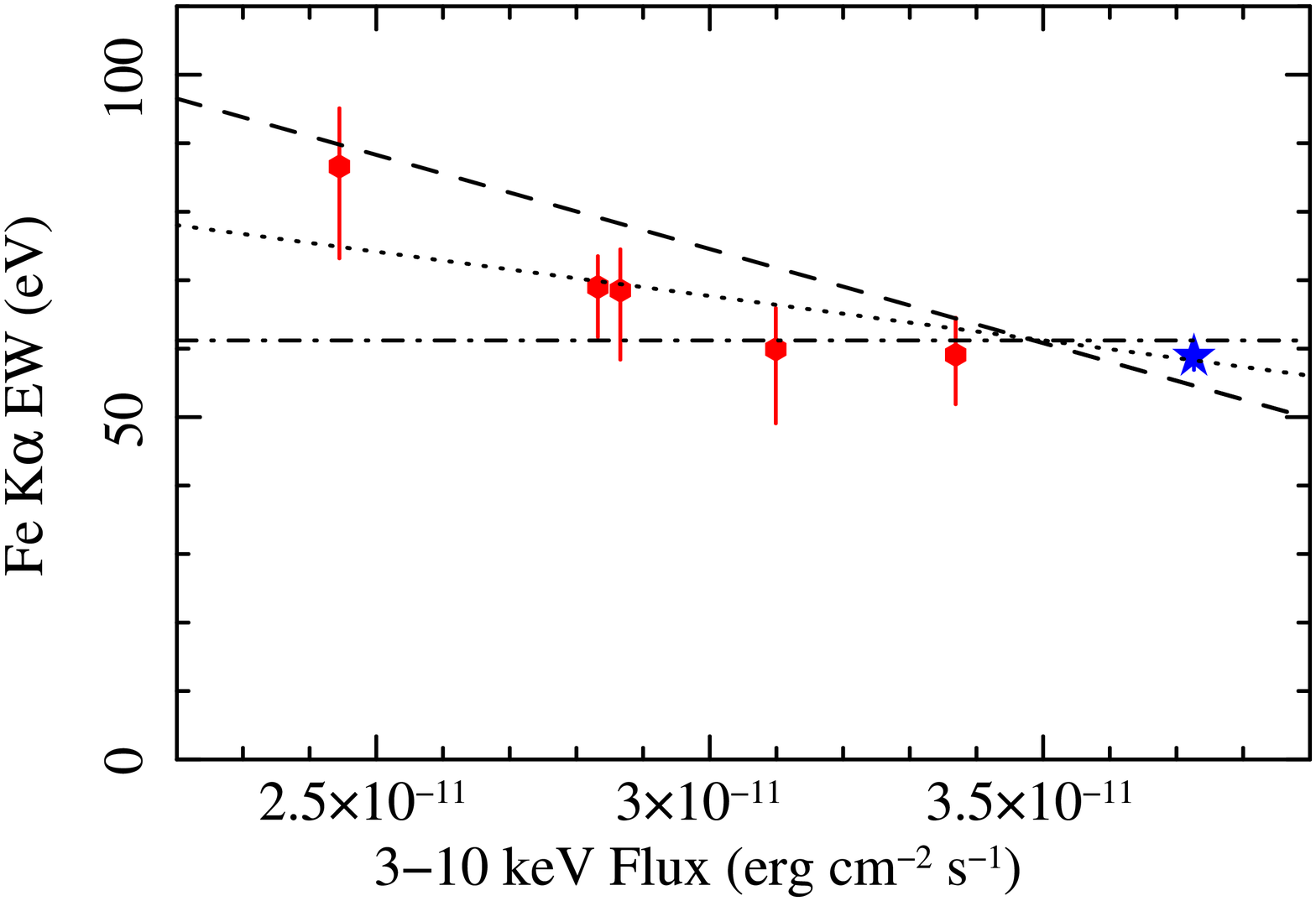}
\caption{{\it (Left, middle and right panels)} Intensity vs. time in MJD, intensity 
vs. 3-10 keV flux and EW vs. 3-10 keV flux of the \FeKa\ line fitted 
with a single Gaussian profile with $\sigma=0.092$ keV (model 1), respectively. 
Each data point represents the best fit result obtained from the fit of the 
spectrum of each \xmm\ observation (red hexagonals) and the total 
spectrum of the 2009 data (blue stars). The dashed lines represent the 
expected relations if the \FeKa\ line has constant intensity, while the 
dash-dotted lines represent a constant EW, and the dotted line shows 
the best fit trend. 
The left panel show the observation date in Modified Julian Date, MJD 
minus 51500 which corresponds to 1999 November $18^{\rm th}$. 
Two \xmm\ observations occur at day 2159 and 2163 and appear 
overlapping in the left panel.}
\label{lcFeK}
\end{figure*}

The upper panel of Fig. \ref{FeKmean} shows, as observed during 
previous observations, an evident emission line at 6.4 keV as well as 
an emission tail at higher energies. 
A simple power law fit to the 4-10 keV total pn spectrum gives an 
un-acceptable fit ($\chi^2=2445.7$ for 1197 dof). 
Adding a Gaussian line to fit the Fe K$\alpha$ line at 6.4 keV improves 
the fit by $\Delta\chi^2=1134$ for 3 extra parameters (see model 1 in 
Tab. \ref{tabMean}).
This line reproduces the bulk of the neutral FeK emission 
(E $=6.43\pm0.01$ keV), but leaves strong residuals at higher energies 
which might be, at least in part, associated with an Fe K$\beta$ emission 
line. We thus add a second line with the same line width as the 
Fe K$\alpha$ line and an energy of 7.06 keV (Kaastra \& Mewe 1993). 
The best fit parameters are given as model 2 in Tab. \ref{tabMean},
and the fit improves by $\Delta\chi^2=66.9$ for the addition of 1 
new parameter.
The best fit intensity is N$_{\rm Fe~K\beta}=0.75\times10^{-5}$ ph cm$^{-2}$ 
s$^{-1}$, and the observed K$\beta$/K$\alpha$ ratio is $=0.19\pm0.02$. 
This is slightly, but significantly, higher than 0.155-0.16, the 
value estimated by Molendi et al. (2003; who assume the Basko 1978 
formulae; but see also Palmeri et al. 2003a,b). The excess of Fe K$\beta$ 
emission might be due to the contamination by the \Fevs\ emission line 
at 6.966 keV. 
This idea is further supported by the observation of clear residuals, 
in the best fit, around 6.7 keV; and by the best fit FeK energy 
($E=6.43\pm0.01$ keV) which is inconsistent with the line arising 
from neutral iron. This thus suggests that the Gaussian line is trying 
to fit both the neutral and ionised Fe K components. 

To model ionised FeK emission, we add two narrow ($\sigma=0$) 
emission lines, one (\Fevc) emitting between 6.637 and 6.7 keV 
(to take into account emission for each component of the triplet)
and the other (\Fevs) emitting at 6.966 keV (model 3 in Tab. \ref{tabMean}). 
Moreover we impose that the intensity 
of the K$\beta$ has to be 0.155-0.16 times the intensity of Fe K$\alpha$ one 
(and $\sigma_{\rm Fe~K\beta}=\sigma_{\rm Fe~K\alpha}$). 
The fit significantly improves ($\Delta\chi^2=25$ for the addition 
of one more parameter). Both \Fevc\ (the best fit line energy is 
consistent with each one of the triplet) and \Fevs\ are statistically 
required (although the \Fevs\ line is not resolved from the Fe K$\beta$ 
emission, thus its intensity depends on the assumed K$\alpha$/K$\beta$ 
ratio). In this model the Fe K$\alpha$ line is roughly consistent with 
being produced by neutral or lowly ionised material 
(E$_{\rm Fe K\alpha}=6.415\pm0.012$ keV). 

Pounds et al. (2001), Ponti et al. (2009), de la Calle et al. (2010), 
Cerruti et al. (2011) and Noda et al. (2011) suggest that the inner 
accretion disc in Mrk509 might be highly ionised and thus the ionised 
emission of the Fe K complex might be associated 
to a relativistic ionised reflection component produced in the inner disc. 
To test this hypothesis, we substitute the two narrow lines (\Fevc\ and \Fevs) 
with one broad ionised line (model 4 in Tab. \ref{tabMean}) with a relativistic 
profile (disc line profile for a Schwarzschild black hole; {\it diskline} 
model in \xspec). We fixed the line energy either to 6.7 or 6.96 keV 
(for the \Fevc\ or \Fevs\ line, respectively) and the inner and outer disc 
radius to 6 and 1000 gravitational radii. In both cases the best fit with 
this model suggests the inner accretion disc to be moderately 
inclined $\sim33-18^{\circ}$, to have a fairly standard disc emissivity 
index $-2.2-2.8$ and an equivalent width of the line to be 
EW$\sim36-40$ eV.
The data are described reasonably well by this model resulting 
in a $\chi^2=1216.0$ and $\chi^2=1219.4$ for 1191 dof, for the case of 
a broad \Fevc\ and \Fevs\ line, respectively. 
Thus, the single broad ionised disc-line model is statistically 
indistinguishable from the multiple narrow lines one ($\chi^2=1219.8$
for 1192 dof). 

\subsection{Medium velocity-high ionisation winds}

The summed spectrum of the previous \xmm\ observations showed 
a medium velocity (v$_{\rm out}\sim0.048\pm0.013$ c) highly ionised 
(Log($\xi$)$\sim5$) outflow in Mrk 509 (Ponti et al. 2009). 
The associated \Fevs\ absorption line was detected both in the EPIC-pn and 
MOS camera, with an equivalent width EW$=-13.1^{+5.9}_{-2.9}$ eV 
and with a total significance between $3-4 \sigma$ ($\sim99.9$ \% probability). 

During the 2009 \xmm\ campaign this highly ionised absorption 
component is not significantly detected. If we add a narrow 
Gaussian absorption line at 7.3 keV, the energy 
of the absorption feature in the previous \xmm\ observations, 
we observe the line to be much weaker with the best fit line EW 
being $-3.2^{+2.6}_{-2.8}$ eV, significantly smaller than observed in 
previous \xmm\ observations. 

\section{The neutral Fe K$\alpha$ component}

In this section we further investigate the nature of the \FeKa\ line, looking 
at the individual spectra obtained over the years. 
Our analysis of the long 2009 monitoring campaign, which triples the total 
exposure on Mrk509, confirms the presence of a resolved component 
($\sigma=0.092\pm0.012$ keV) of the neutral Fe K$\alpha$ line (see model 3, 
but also 4 and 5 of Tab. \ref{tabMean}). 

\subsection{Fe K$\alpha$ variations on years time-scales}

\begin{figure*}
\includegraphics[width=0.46\textwidth,height=0.34\textwidth]{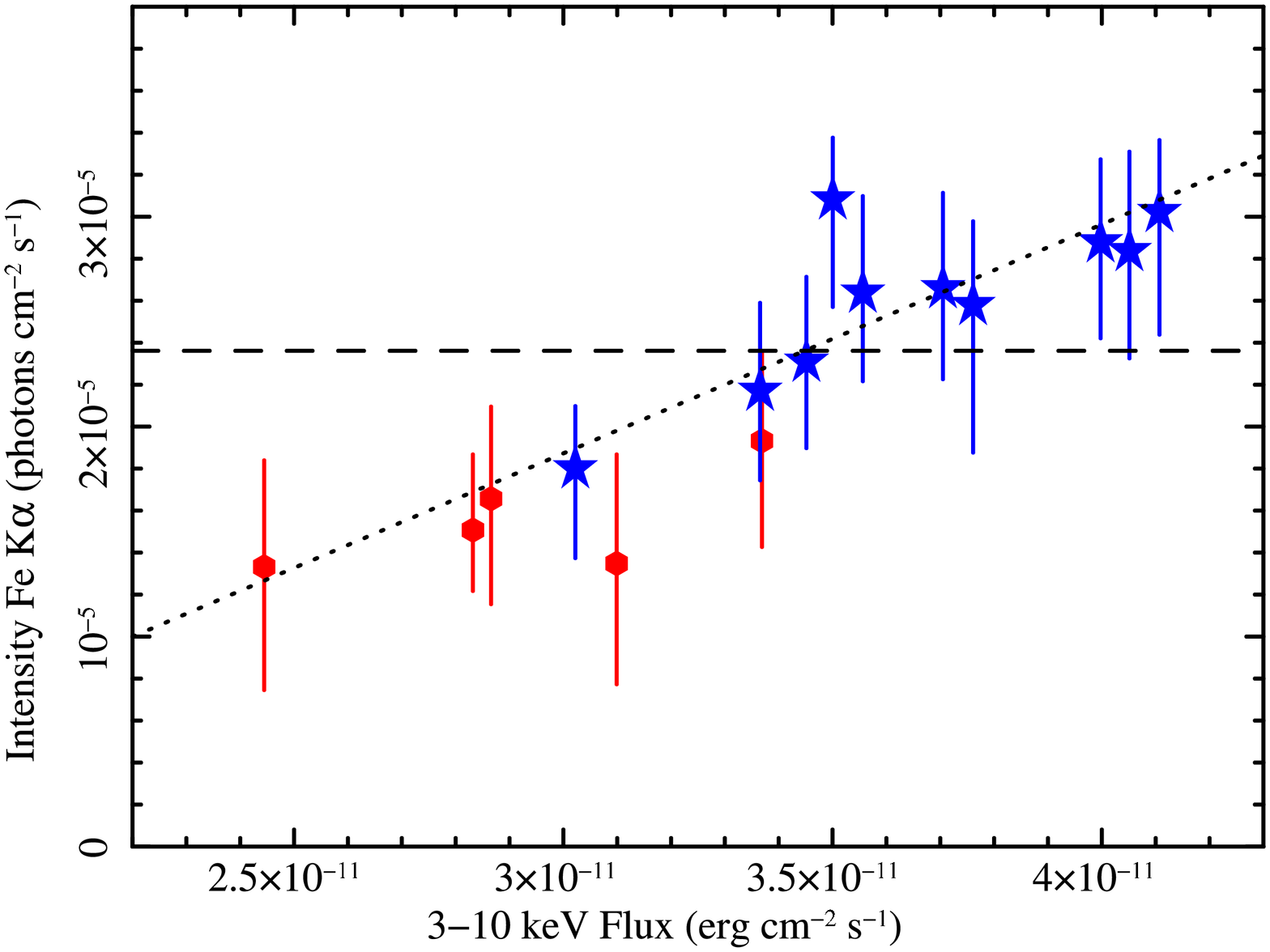}
\includegraphics[width=0.46\textwidth,height=0.34\textwidth]{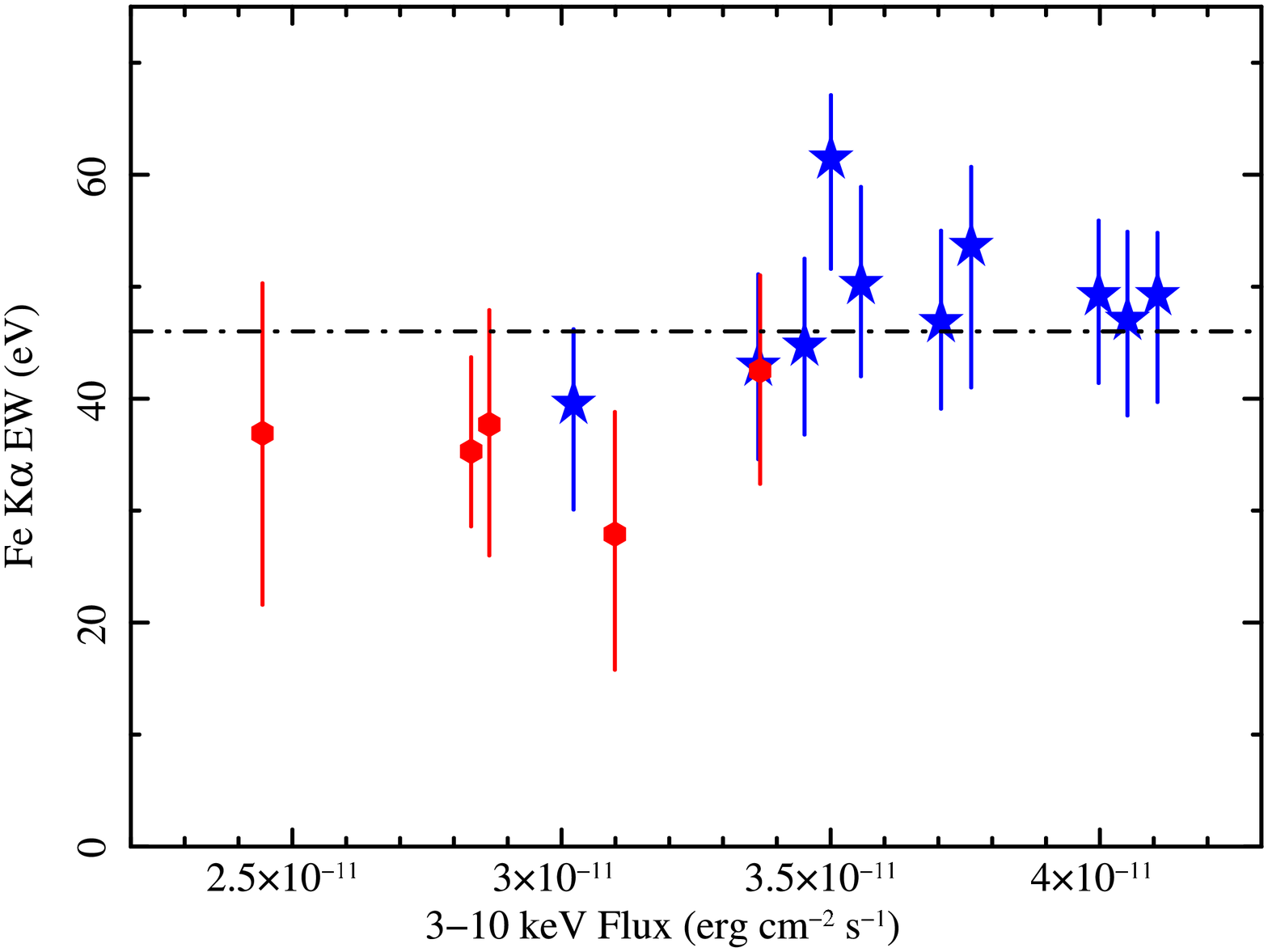}
\caption{{\it (Left and right panel)} The intensity and EW of the resolved 
($\sigma=0.22$ keV) \FeKa\ component (the narrow \FeKa\ component 
is assumed to have $\sigma=0.027$ keV and be constant) as a 
function of the 3-10 keV flux. Blue stars and red hexagonals show the 
best fit results of the 10 observations of the 2009 campaign and of the 
previous \xmm\ observations, respectively. 
Dashed and dot dashed lines show the constant intensity and constant 
EW cases, respectively. 
The best fit relation (dotted line) is consistent with the resolved component 
of the \FeKa\ line having constant EW. The intensity of the resolved 
component of the \FeKa\ line follows the 3-10 keV continuum variations 
with a 1-to-1 relation. 
}
\label{FeKall}
\end{figure*}
To study the variability of the neutral Fe K$\alpha$ emission line (on years 
time-scales), we fit the spectrum of each of the old \xmm\ observations 
(i.e. between 2000 and 2006) with a single FeK$\alpha$ (plus associated 
K$\beta$ emission) plus two narrow emission lines (such as in \S \ref{mean} 
in model 3) to parametrize the ionised FeK emission (see Tab. \ref{tabMean}). 
Leaving the Fe K$\alpha$ widths free to vary, as in Tab. \ref{tabMean}, would 
result in unconstrained values for the spectra with the shortest exposures 
(due to the lower statistics). 
We thus decide to fix the width of the Fe 
K$\alpha$ line to $\sigma=0.092$ keV, its best fit value as observed in 
the total spectrum of the 2009 campaign (model 3 of Tab. \ref{tabMean}). 

The left panel of Fig. \ref{lcFeK} shows the Fe K$\alpha$ line intensity for 
each \xmm\ observation as a function of time, using the summed 2009 
data. The line intensity is observed to vary by less than 25 \%. 
The fit with a constant Fe K$\alpha$ intensity (dashed line), however,
is unsatisfactory ($\chi^2=10.7$ for 5 dof). The middle panel of Fig. 
\ref{lcFeK} shows the Fe K$\alpha$ intensity vs. source flux in the 3-10 keV band. 
The fit slightly improves when a linear relation (see dotted line) is considered 
($\Delta\chi^2=7.6$ for the addition of one more parameter; 97 \% F-test probability). 
The increase in Fe K$\alpha$ intensity with flux might suggest that the line 
is responding quickly to the illuminating continuum, keeping a constant 
EW with flux. The dash-dotted line shows the expected Fe K$\alpha$ intensity 
variation for a line with constant EW. The observed line intensity variations are 
intermediate between the constant intensity and constant EW cases. 
The right panel of Fig. \ref{lcFeK} confirms that the line has neither a constant 
intensity nor constant EW, instead it sits somewhere in the middle between these 
two cases. 

The \FeKa\ variations on years time-scales suggest that at least part of the 
line is varying following the 3-10 keV continuum. 
We want to point out that the width of the Fe K$\alpha$ line is comparable 
to the EPIC-pn energy resolution. This means that the observed \FeKa\ variability 
may be the product of a constant narrow component, coming from distant material,
plus a broader, resolved and variable \FeKa\ line produced closer to the BH. 
Unfortunately, due to the limited energy resolution of the EPIC cameras aboard 
\xmm, we cannot resolve the Fe K emission, coming from regions located 
at light weeks from those at light years from the BH, based 
on the \FeKa\ line widths. Only with the \chandra\ high energy 
transmission grating (HETG) we can confidently pose some constraints 
on the distance of the different Fe K emission components. 

\subsection{\chandra\ HETG}

\chandra\ observed Mrk 509 with the HETG instrument only once for 
50 ks (Yaqoob et al. 2003).
During the HETG observation the 3-10 keV flux was $4.36\times
10^{-11}$ erg cm$^{-2}$ s$^{-1}$ with a power law spectrum of  
index $\Gamma=1.76^{+0.03}_{-0.02}$. An excess was present at 6.4 keV, 
thus we added a Gaussian line\footnote{In order not to lose the 
excellent energy resolution, we decided not to rebin the spectrum and to fit 
the data using the C-statistics (Cash 1979). The best fit has C-stat value 
of 370.5 for 284 dof.}. In agreement with the results obtained 
by Shu, Yaqoob \& Wang (2010) and Yaqoob \& Padmanabhan (2004) we 
detect a line at $6.42\pm0.02$ keV with an intensity of 
$3\pm2\times10^{-5}$ photons cm$^{-2}$ s$^{-1}$. 
The line is resolved and has a width significantly smaller than the one 
measured by \xmm, $\sigma=0.027^{+0.018}_{-0.010}$ keV. 
This suggests that, at least part of, the neutral Fe K$\alpha$ emission 
is produced in regions more distant than a few thousands gravitational radii.
In fact, if the material is in Keplerian motion and assuming a BH mass of Mrk 509 
of M$_{BH}=1.4-3\times10^8$ M$_\odot$ (Peterson et al. 2004; Mehdipour et al. 2011), 
then the narrow core of the line is produced at a distance of r=0.2-0.5 pc ($\sim30000$ r$_g$). 
We note that this value is of the same order of magnitude 
as the molecular sublimation radius for Mrk 509. Landt et al. (2011) using 
quasi-simultaneous near-infrared and optical spectroscopy estimate 
a radius of the hot dust of $\sim0.27$ pc (0.84 ly), which is also consistent 
with the one estimated following eq. 5 of Barvainis (1987) assuming a bolometric 
luminosity L$_{\rm Bol}=1.07\times10^{45}$ erg s$^{-1}$ (Woo \& Urry 2002). 
This suggests that this narrow component of the Fe K line might be associated 
to the inner wall of the molecular torus.

\subsection{Two components of the \FeKa\ line}
\label{twolines}

Thus, as suggested by the analysis of the \chandra\ HETG data and in 
agreement with the observed variability on years time-scales, we 
interpret the Fe K$\alpha$ line as being composed by two components 
which are indistinguishable at the EPIC resolution. 
We first re-fit the mean spectrum of the 2009 campaign 
with two components for the Fe K$\alpha$ line. One "narrow" component
with the line width fixed at the best fit value derived from \chandra\ 
analysis ($\sigma=0.027$ keV) plus a "resolved" component 
with its width free to vary. 
Model 6 in Tab. \ref{2lines1} shows the best fit results assuming 
the energy of both components is the same.
The narrow component has a best fit EW$=27\pm4$ eV, 
while the resolved neutral line has an EW$=42^{+9}_{-4}$ eV
and a best fit line width $\sigma=0.22\pm0.05$ keV which is larger than 
the single Gaussian \FeKa\ fit ($\sigma=0.092$ keV). 

Reflection is the most probable origin of the Fe K$\alpha$ line. 
Associated to reflection lines an underlying reflection continuum is 
expected and generally observed. In particular the ratio between 
the intensity of the line over the reflection continuum strongly 
depends on the reflector column density reaching a value of 
EW$_{\rm Fe K\alpha}\sim 1$ keV for Compton thick materials. 
To check the impact of the reflection continuum on the best fit 
model we add a standard neutral reflection continuum ({\sc pexrav} 
in {\sc Xspec}) with intensity such that the EW$_{\rm Fe K\alpha}=1$ 
keV over their reflection continua. The new best fit line EW do not 
vary significantly (EW$=26\pm3$ eV and EW$=33^{+8}_{-5}$ eV, 
for the narrow and resolved Fe K$\alpha$ lines, respectively). 
Thus, and considering also the limited energy band used here 
we decide to disregard the continuum reflection component
(the impact of the reflection component on the broad band 
source emission will be studied by Petrucci et al. 2012 
taking into account the UV to soft Gamma ray emission with 
physical models).
The narrow component, as observed during the 2009 campaign, has an 
intensity of $1.5\pm0.2\times10^{-5}$ ph cm$^{-2}$ s$^{-1}$, which is about 
half the total \FeKa\ intensity (see Tab. \ref{tabMean} and Fig. \ref{lcFeK}). 
This value is consistent with the intensity of the narrow \FeKa\ line 
observed by \chandra.
The width of the narrow component suggests a distance of 0.2-0.5 pc from 
the BH, thus we expect that all the variability on time-scales shorter 
than a few years would be smeared out because of light travelling effects. 
For this reason and because the lower signal to noise in the individual 2009 
and earlier \xmm\ spectra does not allow us to disentangle both 
components, in all the following fits we assume a constant intensity 
of $1.5\times10^{-5}$ cm$^{-2}$ s$^{-1}$ for this narrow \FeKa\ component. 

Next, we fit the spectrum of each of the 15 \xmm\ observations 
with model 6 shown in Tab. \ref{2lines1}, assuming a constant intensity for 
the narrow \FeKa\ component and constant width for the resolved 
component, plus the associated Fe K$\beta$ and narrow \Fevc\ and \Fevs\ lines.
The left panel of Fig. \ref{FeKall} shows the intensity of the resolved 
component ($\sigma=0.22$ keV) vs. the 3-10 keV flux for the 10 
observations of the 2009 monitoring (blue stars) as well as for the 
previous \xmm\ observations (red hexagonals). 
The dashed line shows the best fit assuming that the \FeKa\ intensity is constant,
which results in an unsatisfactory fit with $\chi^2=33.2$ for 14 dof. 
On the other hand, once the data are fitted 
with a linear relation (dotted line), the fit significantly improves 
($\Delta\chi^2=23.0$ for the addition of 1 new parameter, which 
corresponds to F-test probability $>99.98$ \%). 
We also compute the Pearson's linear correlation coefficient is equal 
to 0.87 and has a probability $=2.7\times10^{-5}$, which corresponds to a 
significance of the correlation of more than 4 $\sigma$ (similar results 
are obtained using a Spearman's rho or Kendall's tau correlation coefficients). 
The slope of the observed best fit relation is consistent with that expected 
if the resolved line is responding linearly to the continuum variations. 
This is confirmed in the right panel of Fig. \ref{FeKall} which shows the 
resolved \FeKa\ line EW is consistent with being constant ($\chi^2=17.1$ 
for 14 dof), as expected if the \FeKa\ line is responding linearly to the 3-10 
keV continuum flux variations.

The line intensity is significantly variable even on time-scales of few days, 
e.g. between the different pointing of the 2009 monitoring campaign. In fact, 
fitting the 2009 Fe K$\alpha$ intensities with a constant gives a 
$\chi^2=13.7$ for 9 dof, which becomes $\chi^2=5.3$ when a linear 
relation is considered (as specified in \S 2, conservative 90 \% errors 
are used here). The Pearson's linear correlation coefficient results to be 
0.8 and has a probability $=5\times10^{-3}$, which corresponds to a significance 
of the correlation of about 3 $\sigma$.

\begin{figure}
\includegraphics[width=0.47\textwidth,height=0.32\textwidth,angle=0]{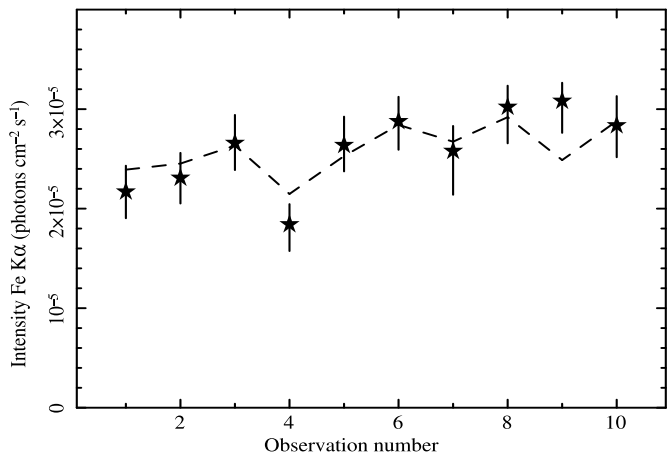}
\caption{Intensity vs. observation number of the 10 \xmm\ pointings of the 
2009 campaign. The black dashed line shows the re-scaled (with the mean 3-10 keV 
flux equalling the mean line intensity) 3-10 keV source flux. The intensity of the 
resolved component of the \FeKa\ line follows, with a 1-to-1 relation the 3-10 keV 
continuum variations without any measurable lag. $1 \sigma$ errors are shown.} 
\label{varbroad}
\end{figure}
Fig. \ref{varbroad} shows the variations of the intensity of the resolved 
component of the \FeKa\ line as a function of time (1 $\sigma$ errors are shown
here), during the 2009 
campaign, overplotted on the 3-10 keV rescaled flux (dashed line). 
As already suggested above and Fig. \ref{FeKall}, 
the Fe K line variations track that of the continuum very well.
We also note that no measurable lag is present, thus this broad 
component of the \FeKa\ line responds to the X-ray continuum within 
less than four days. 

\subsection{Total RMS spectra}

\begin{figure}
\includegraphics[width=0.49\textwidth,height=0.38\textwidth,angle=0]{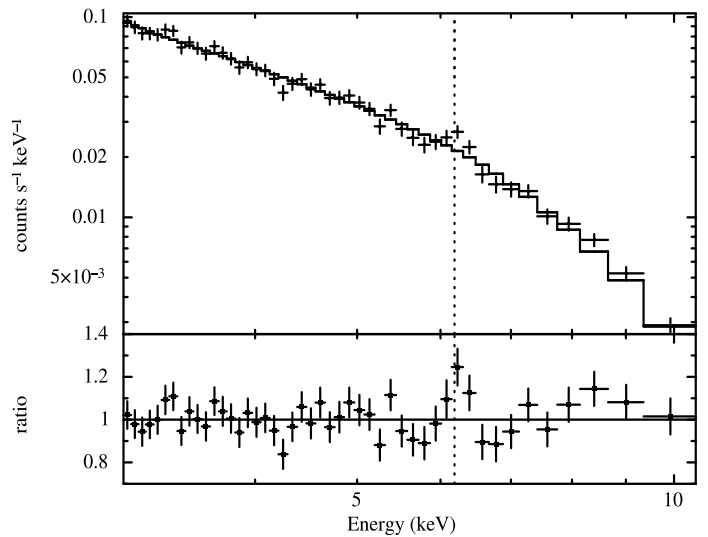}
\caption{Long time-scales total rms spectrum between the different observations. 
The spectrum is calculated with time bins of 60 ks (corresponding to the exposure 
of each \xmm\ pointing) and the total monitoring time of about 1 month. 
An excess of variability is clearly evident at 6.4 keV (EW$=71\pm36$ eV). 
This excess confirms, in a model independent way, the correlated variability of 
the resolved Fe K line component (see Fig. \ref{varbroad}).
}
\label{RMS}
\end{figure}
Figure \ref{RMS} shows the total Root Mean Square variability (rms) 
spectrum calculated between the 10 different observations of the 2009 
campaign. The rms has been calculated with ten 
time bins, each one being a 60 ks \xmm\ pointing. Thus this rms is sampling 
the variability within the observation separation time-scale of about 4 days and 
the monitoring time-scale of slightly more than one month (see Fig. \ref{lc}).
The total rms shows the spectrum of the variable component, only. Thus, 
in contrast to the mean spectrum, it has no contribution from the constant 
emission from distant material (i.e. the narrow core of the Fe K line).
The uncertainties on the total rms are derived from the uncertainties on 
the fractional variability (see formula B.2 of Vaughan et al. 2003; A.1 of 
Ponti et al. 2004) multiplying for the mean and taking into account its error.

The 3-10 keV total rms spectrum has a power law shape with spectral index 
$\Gamma=1.98\pm0.06$ and normalisation of $1.05\pm0.05\times10^{-3}$ 
ph. cm$^{-2}$ s$^{-1}$ (see Fig. \ref{RMS}). A clear excess 
of variability is present at 6.4 keV. The addition of a Gaussian line significantly 
improves the fit ($\Delta\chi^2=10.6$ for the addition of 2 dof, that corresponds 
to a F-test significance of 99.3 \%). The best fit energy of the line is 
$E=6.45\pm0.08$ keV and intensity $(2\pm1)\times10^{-6}$ ph. cm$^{-2}$ 
s$^{-1}$. 
The width of the line is constrained to be less than $\sim0.2$ keV,  
consistent with the variable resolved neutral Fe K line. The equivalent 
width in the total rms spectrum is $EW=71\pm36$ eV consistent 
with the EW of the resolved component observed in the mean spectrum. 
The detection of this excess of variability indicates, in a model independent 
way, that the resolved component of the line is varying linearly with the 
continuum on these time-scales. This reinforces the robustness of the 
detection of Fe K reverberation.

\subsection{Locating the \FeKa\ emitting region}

Are the observed variability properties in agreement with the spectral ones?
Assuming that the material producing the resolved Fe K line is in Keplerian 
motion around the BH, the line width ($\sigma=0.22$ keV) implies that it is located 
at $300-1000$ r$_g$ from the BH. Assuming a BH mass of Mrk 509 
of M$_{BH}=1.4-3\times10^8$ M$_\odot$ (Peterson et al. 2004; Mehdipour
et al. 2011), then this distance corresponds to about few days - a light week. 
The typical spacing between the different \xmm\ observations during the 2009 monitoring 
is about 4 days, thus it is in very good agreement with the observed \FeKa\ variations.
Moreover, the fast response of the \FeKa\ flux to the continuum changes 
indicates that the bulk of the resolved \FeKa\ emission is produced 
around or within several hundreds up to few thousands gravitational radii from 
the central BH. 

It is more difficult, however, to pose a lower limit to the position of the \FeKa\ emitting 
region. We note that if the \FeKa\ line emitting region extends down 
to few gravitational radii from the BH, then the line shape should present 
a prominent red wing. Thus, we fit the resolved component of the \FeKa\ line 
with a disc line profile ({\sc diskline} in {\it Xspec}). In the fitting process we allow 
the line energy to vary in the range 6.4-6.42 keV, we fix the disc outer radius to 
1000 r$_g$ and the illumination profile to $\alpha=-3$ (the expected value for a 
standard alpha disc; Laor 1991; Wilkins \& Fabian 2011). 
The lack of relativistic redshifted \FeKa\ emission 
suggests an inner radius larger than $\sim85$, 45 and 37 r$_{\rm g}$
(which corresponds to roughly 8-16 light hours) for a disc inclination of 
30, 20 and 10 degrees, respectively. 
Thus, the resolved component of the \FeKa\ line is probably emitted between 
$40-1000$ r$_g$ from the BH. 

Can such a "narrow" disc annulus produce an \FeKa\ line of $\sim40-50$ eV? 
Reflection from an accretion disc with Solar iron abundances covering half 
of the sky is expected to produce a \FeKa\ line with EW$\sim100-150$ eV
(Matt et al. 1991). The line EW is expected to decrease/increase roughly 
linearly/logarithmically for iron abundances lower/higher than Solar (Matt et al. 
1996; 1997). Steenbrugge et al. (2011) measured in Mrk 509 a relative iron 
to oxygen abundance of Fe/O$=0.85\pm0.06$. Assuming that this translate 
into an iron abundance of 0.85 Solar (but see Arav et al. 2007), this would 
correspond to EW$_{\rm Fe K\alpha}\sim90-130$ eV. 
If the primary X-ray source in Mrk 509 is compact (as the variability 
suggests; McHardy et al. 2006; Ponti et al. 2012) and located at a 
few r$_g$ above the BH and if the disc is flat, we can estimate (neglecting 
relativistic effects) the geometric solid angle covered by the disc annulus 
producing the 
\FeKa\ line (r$_{\rm in} \sim 40$ and r$_{\rm out}\sim1000$ r$_g$). 
If the primary X-ray source is located between 1 and 4 r$_g$ above the BH 
(De Marco et al. 2012), the flat disc annulus covering factor would be between 
2 and 5 \% of the sky. Thus reflection from such a flat annulus would produce 
(even in the extreme case of a \FeKa\ EW$=130$ eV for a standard disc) 
a line with EW$\sim4-13$ eV. 
The observed EW of the resolved-variable \FeKa\ line is several times 
larger (EW=42 eV) than this estimated value. This indicates a larger covering 
factor of the reflector, compared to the flat disc, suggesting that the material 
producing the \FeKa\ line is distributed azimuthally above the disc, possibly 
in the form of clouds, perhaps associated to the inner BLR (see Costantini 
et al. 2012 for more details). 

The observed correlation on days-weeks timescales constraints also 
in which part of the BLR the Fe K$\alpha$ line is produced.
We can exclude, in fact, that the \FeKa\ emission is produced in the 
optical BLR (producing the bulk of H$\beta$ emission), because the 
H$\beta$ line is observed to reverberate with a delay of 80 days, 
being thus significantly more distant than the region producing 
the \FeKa\ line. On the other hand, several studies show that the 
BLR might be stratified, with the higher ionisation lines 
located closer to the central BH. The correspondence between \FeKa\ and 
the inner BLR is reinforced by the consistency between the width of the \FeKa\ 
line ($\sigma=0.21\pm0.07$ keV, which corresponds to 
$FWHM\sim1.5$$-$$3 \times10^4$ km s$^{-1}$) and the ones of the 
broadest components of the UV broad emission lines (e.g. Ly$\alpha$, 
C {\sc iv}, C {\sc iii} and O {\sc vi}) which have  
components with $FWHM\sim10^4$ km s$^{-1}$ (Kriss et al. 2011). 

The lack of relativistic effects on the shape of the \FeKa\ line suggests the 
absence of a neutral standard thin accretion disc extending down to a few 
gravitational radii from the BH. However, this appears to be at odds with the 
high efficiency (L$_{\rm Bol}\sim5-10$ \% L$_{\rm Edd}$) of the disc emission 
of Mrk~509 (Mehdipour et al. 2011; Petrucci et al. 2012). 
This leads to the question of why we do not find traces of the inner accretion 
disc in the \FeKa\ line shape if it is present in this source. 

\section{The ionised Fe K emission}

During both the 2009 campaign and the previous \xmm\ observations, 
Mrk 509 clearly showed an excess of emission around 6.7-7 keV 
(see Fig. \ref{FeKmean}) most probably associated to emission 
from ionised iron. As shown in \S \ref{mean} this excess can be modelled 
both by the combination of narrow emission lines from \Fevc\ and \Fevs, 
or by a single relativistic emission line (see Tab. \ref{tabMean}). 
The parameters of this weak ionised emission line(s) can be affected by 
the modelling of the stronger Fe K$\alpha$ line. For this reason, 
we now re-fit the mean spectrum including both the narrow and the 
resolved component of the Fe K$\alpha$ line. 

We first consider that the ionised emission is produced by narrow 
emission lines (\Fevc\ and \Fevs). Such emission lines from highly 
ionised ions are now often observed (Costantini et al. 2010; e.g. 
for a compilation of sources, Bianchi et al. 2009a,b; Fukazawa et al. 2011) 
and they can arise from photo-ionised (Bianchi et al. 2005; 
Bianchi \& Matt 2002) or collisionally ionised plasma (Cappi et al. 1999). 
Thus we fit the spectrum with two components for the \FeKa\ line and 
the associated K$\beta$ lines, plus two narrow ($\sigma=1$ eV) Gaussian 
emission lines one (\Fevc) with energy constrained to be between $6.637$ 
and $6.7$ keV and the other (\Fevs) with energy fixed at $E=6.966$ keV (see 
model 6 in Tab. \ref{2lines1}). 
The model reproduces the data well ($\chi^2=1198.5$ for 1190 dof). 
The weakness of the \Fevc\ and \Fevs\ lines prevents us from significantly 
constraining the line variability between the different \xmm\ observations. 

For comparison, we fit with the same model also to the summed 
spectrum of all \xmm\ observations taken between 2000 and 2006 
(see model 7 of Tab. \ref{2lines1}). 
The ionised emission lines are consistent with being constant within 
the two sets of observations. However, the statistics is not good 
enough to discriminate if it is the line intensity (which would suggest 
an origin at large distances) or the EWs remain constant. 
Clear is, instead, the variation of the medium outflow velocity highly 
ionised absorption line, which almost disappeared during the 2009 
campaign. 
The addition of this component to the model used to fit the combined spectra 
of the 2009 campaign, improves the fit by $\Delta\chi^2=4.5$ for 2 
new parameters which corresponds to a F-test probability of $\sim90$ \%. 
Thus we decide to disregard this absorption component in all subsequent fits.

Another clear difference compared to previous observations is 
related to the disappearance of the highly ionised absorption with 
mildly relativistic (up to 0.14-0.2c) outflow velocities (Dadina et al. 2005; 
Cappi et al. 2009; Tombesi et al. 2010). 
We searched, in fact, for such features in all 10 observations obtained after the 
\xmm\ campaign, by including narrow absorption lines in the model between 4-10 keV. 
We found only a marginal ($\Delta \chi^2$ $\sim$6) detection of two absorption 
features at 9 keV and 10.2 keV (rest-frame energies) during observation 4.
Even if consistent with being produced by \Fevs\ K$\alpha$ and K$\beta$ at 
v$\sim$0.3c, and similar to earlier results (Cappi et al. 2009), the level of (highly) 
ionised absorption during the \xmm\ campaign is found to be 
significantly reduced compared with most previous \xmm\ observations. 
We obtained upper limits (at 90 \% confidence) on the equivalent width of 
narrow ($\sigma$ fixed to 100 eV) Gaussian absorption lines with typical 
values between -5 and -30 eV, between 7.5 and 9.0 keV, depending on the 
energy and observation considered. This is typically lower than values (between 
-20 and -30 eV) found in the lines detected in earlier observations (Cappi et al. 
2009, Tombesi et al. 2010) excluding that such UFOs were present during the 
2009 campaign.

\subsection{Collisionally ionised plasma}

We attempt to interpret the highly ionised emission lines via a 
self-consistent physical model. 
First we applied the collisionally ionised model 
CIE\footnote{http://www.sron.nl/files/HEA/SPEX/manuals/manual.pdf}  
in spex (Kaastra et al. 1996). 
In this fit we considered the 3.5--10\,keV band for the continuum. 
We used Gaussian components for the \FeKa\ line profile and Fe K$\beta$, 
constraining the flux of the latter to be 0.155-0.16 times the \FeKa\ one (Palmeri et al. 2003). 
This was done in order to mitigate the degeneracy induced by the partial 
blend with the \Fevs\ Ly\,$\alpha$ line. 
The best fit points to a high-temperature gas ($k$T$=8.5\pm1.5$\,keV). At this temperature,
the predicted line fluxes are $\sim5\times10^{-6}$\,ph\,cm$^{-2}$\,s$^{-1}$ and 
$\sim3\times10^{-6}$\,ph\,cm$^{-2}$\,s$^{-1}$ for the \ion{Fe}{xxv} triplet and the 
\ion{Fe}{xxvi}\,Ly\,$\alpha$, respectively. 
These values are consistent with those measured empirically using Gaussian 
lines (i.e. Table~\ref{2lines1}). In theory these lines may be produced by hot, line 
emitting, gas in the form of a starburst driven wind. Mrk 509 has a total 
luminosity L$_{\rm 2-10~keV}=1.3\times10^{44}$ erg s$^{-1}$ in the 2-10 
keV band. Assuming an Fe abundance of 0.4 Solar (as observed in 
starburst galaxies, Cappi et al. 1999) the best fit thermal starburst model 
requires a luminosity L$_{\rm 2-10~keV}=3.3\times10^{42}$ erg s$^{-1}$
to reproduce the \Fevc\ and \Fevs\ line emission (reducing to 
L$_{\rm 2-10~keV}=1.6\times10^{42}$ erg s$^{-1}$ for Solar iron 
abundance). 
Using the correlation between L$_{\rm 2-10~keV}$
and the far infrared luminosity (L$_{\rm FIR}$), valid in star forming galaxies 
(Ranalli et al. 2003), we estimate a corresponding L$_{\rm FIR}>10^{46}$ erg 
s$^{-1}$ and a star formation rate higher than 400 M$_{\odot}$ yr$^{-1}$, 
which is several times larger than the actual total IR luminosity of Mrk 509, 
L$_{\rm IR}\sim2\times10^{11}$ L$_{\odot}$ $\sim8\times10^{44}$ erg s$^{-1}$ 
(Rieke et al. 1978), thus we disfavour this interpretation.

\subsection{Photo-ionised plasma}

\begin{figure}
\begin{center}
\includegraphics[width=0.34\textwidth,height=0.46\textwidth,angle=90]{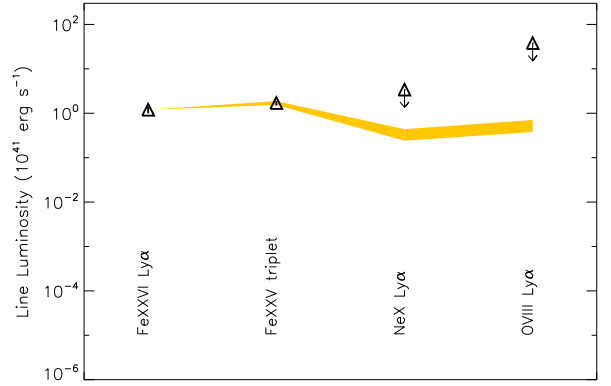}
\end{center}
\caption{\label{f:high_ion} Photoionization modeling for the highly 
ionized iron ions \ion{Fe}{xxvi} and \ion{Fe}{xxv}. Triangles: data. 
Light shaded line: range of models viable to fit the data.}
\end{figure}

Alternatively, the highly ionised lines may be produced by a 
photoionised plasma. 
To test this we used a grid of parameters created using Cloudy 
\citep{ferland} where the column density log$N_{\rm H}$ of the gas 
ranged between $21.7-24.5$\,cm$^{-2}$ and the ionisation parameter 
log($\xi$) ranged between $3.4-7$. The grid has been calculated 
using a covering factor of one. Since the intrinsic line luminosity scales 
linearly with the covering factor, we used the ratio between the model and 
the data of the \ion{Fe}{xxvi} line as a reference for the covering factor. 
In Mrk 509, only  \ion{Fe}{xxvi} and \ion{Fe}{xxv} have significant detection, 
while for other narrow lines from highly ionised ions (e.g. \ion{O}{viii} 
at 18.97 $\AA$ and \ion{Ne}{X} at 12.13 $\AA$) we have obtained upper limits 
from the RGS spectrum. These limits are useful in constraining the model 
(i.e. Costantini et al. 2010). 
In Fig.~\ref{f:high_ion} we compare the line luminosities observed with those 
computed for a range of models which can fit the data. 
In order to reproduce the luminosity of the highly ionised iron ions, the gas 
should have log($\xi$)$=4-5.1$ and $N_{\rm H}=23.4-24.2$. The covering factor 
is $C_V=0.3-0.5$. As we do not see any associated absorption, the gas 
must be out of the line of sight. 
Such lines might possibly originate in e.g. the narrow line region or the highly 
ionised skin of the torus (Bianchi et al. 2005; Bianchi \& Matt 2002).

\subsection{Ionised reflection from inner disc}

As shown in \S \ref{mean}, the ionised emission can be fitted equally 
well with an ionised relativistic emission line. A comparably good fit is 
also obtained with a broad Gaussian \Fevc\ profile in addition 
to the double Fe K$\alpha+\beta$ lines. 
We fix the energy of the broad \Fevc\ line to E$=6.7$ keV. 
The line is significantly broadened $\sigma=0.23$ keV and 
moderately intense (EW$=15$ eV). 
Although this broadening is not as extreme as to exclude a simple Compton 
broadening on the ionised surface of the accretion disc, we decided,
as an alternative to the Gaussian line, to fit the ionised emission with a 
relativistic disc line profile. 
An acceptable fit is obtained also in this case ($\chi^2=1204.7$ for 
1192 dof). Assuming a standard disc emissivity index, inclination and 
outer disc radius of $\beta=-3$, $\alpha=30^{\circ}$ and 
r$_{\rm out}=200$ r$_g$, the best fit disc inner radius is 
r$_{\rm in}=27$ r$_g$, consistent with a value as small as 
r$_{\rm in}=7$ r$_g$ and the disc-line equivalent width is, 
EW $=20^{+6}_{-8}$ eV. 

It is difficult to estimate the line EW expected from an ionised inner 
annulus of the disc. In fact the \Fevc\ and \Fevs\ line EW strongly 
depend on the poorly constrained disc ionisation parameter (Garcia 
et al. 2011) and on the disc annulus covering angle. However for 
reasonable values of these parameters (annulus covering angle 
$\sim 20-40$ \% of the sky and log($\xi$)$\sim2.8-3.5$ erg cm$^{-2}$ 
s$^{-1}$, which corresponds to the peak of \Fevc\ and \Fevs\ emission) 
the line EW is expected to be between 5-50 eV. These results 
suggest that the inner part of the accretion disc might be 
highly ionised, thus explaining the lack of detection of relativistic 
\FeKa\ line, combined with the high source efficiency (which suggests 
a thin standard accretion disc extending all the way down to the last stable orbit). 

\section{Discussion and Conclusions}

We investigated the spectral variability of the Fe K band in the nearby, 
bright Seyfert 1 galaxy Mrk 509, using the 10 observations of the 2009 
\xmm\ monitoring campaign as well as all the previous \xmm\ observations, 
totalling an exposure of more than 900 ks in about 10 years, 
resulting in one of the best quality Fe K spectra ever taken of a 
Seyfert 1 galaxy. This allows us, for the first time, to perform 
reverberation mapping of the resolved Fe K$\alpha$ line. 

\begin{figure}
\begin{center}
\includegraphics[width=0.42\textwidth,height=0.52\textwidth,angle=-90]{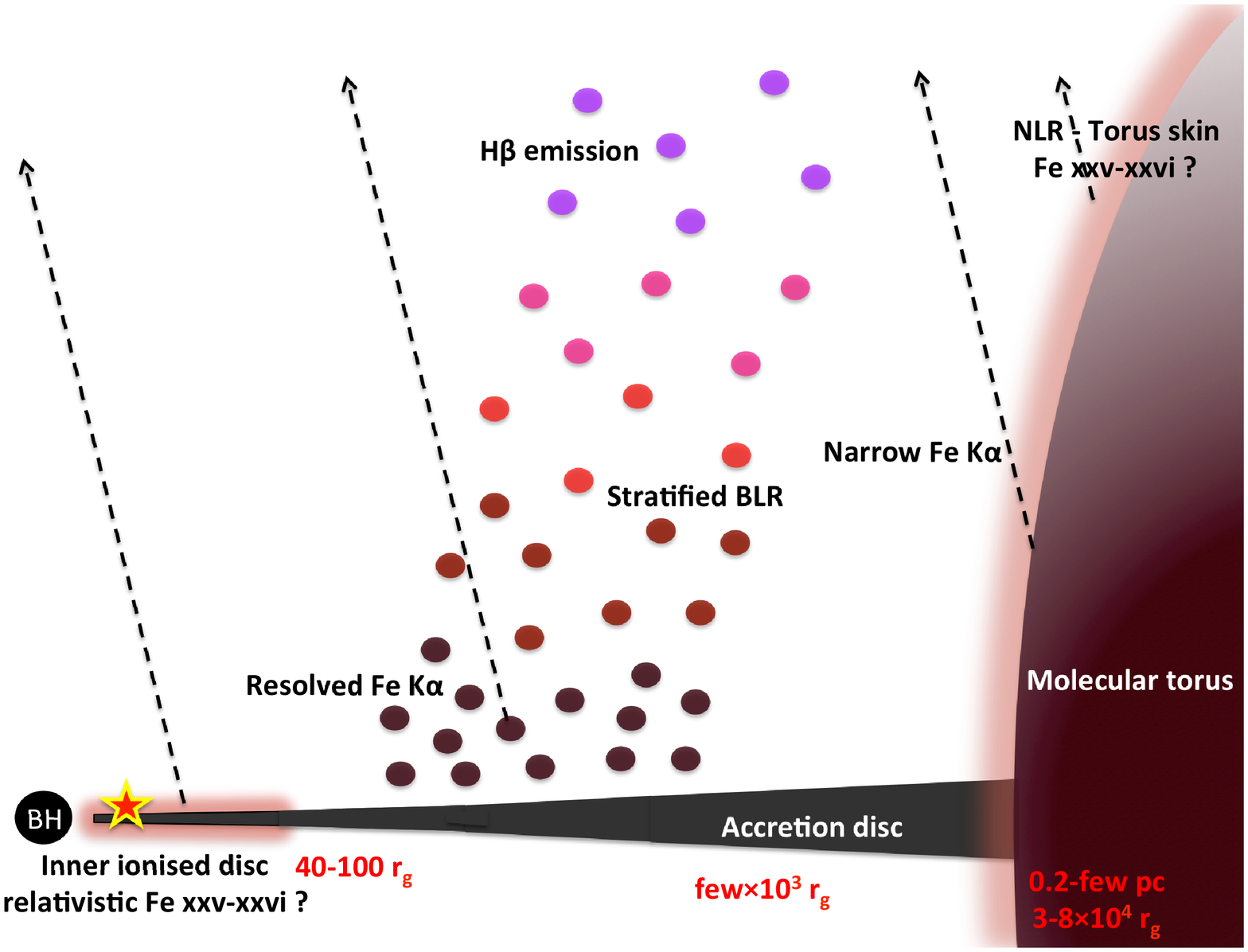}
\end{center}
\caption{Sketch of a possible locations for the different regions 
producing Fe K emission (diagram not to scale). The star represents
the primary X-ray source, located close to the BH.} 
\label{sketch}
\end{figure}

Figure \ref{sketch} summarise in a sketch a possible scenario for 
the production of the Fe K emission in Mrk 509. 
The width of the narrow core of the Fe K$\alpha$ line suggests 
an origin from distant material, possibly the inner wall of the 
molecular torus located at 0.2-few pc. The correlated 
variations (on a few days time-scales) between the 3-10 keV 
continuum and the intensity of the resolved component of the 
Fe K$\alpha$ suggest an origin between several tens and 
few thousands of r$_{\rm g}$ from the BH. The resolved 
Fe K$\alpha$ emission can be produced in the disc, but 
we favour an origin at the base of a stratified broad line region.
We note that none of the X-ray or UV absorption components 
with measured location is co-spatial with the resolved Fe K$\alpha$ 
emitting region. Moreover, the properties of the X-ray and UV 
absorbers appear to differ from the ones required to produce 
the resolved Fe K$\alpha$ line, suggesting that this emitting 
material is outside the line of sight, possibly in the form of 
an equatorial disc flattened wind such as observed in stellar 
mass black holes in the soft state (Ponti et. al. 2012) and 
neutron stars (Diaz-Trigo et al. 2006).
The ionised Fe K emission might be produced either by 
photo-ionisation from distant material, such as the narrow line 
region and/or the ionised skin of the torus, or in the ionised 
inner accretion disc.

The results of this study show that:

\begin{itemize}

\item{} The \xmm\ spectrum of Mrk 509 shows an evident \FeKa\ line 
with total EW$=58\pm4$ eV. Fitted with a single Gaussian line 
the width is $\sigma=0.092\pm0.012$ keV. 
The line intensity increases with the 3-10 keV flux, but not as strongly
as expected in a constant EW scenario, suggesting the presence of 
a constant and a variable \FeKa\ line component. 

\item{} The \chandra\ HETG spectrum has enough energy resolution 
to resolve the narrow component of the \FeKa\ line 
($\sigma=0.027^{+0.018}_{-0.010}$ keV; line intensity 
($1.5\pm0.2$)$\times10^{-5}$ ph cm$^{-2}$ s$^{-1}$). 
The width of the narrow component of the line suggests an origin 
at around 0.2-0.5 pc ($\sim30000$ r$_{\rm g}$) from the BH. This value 
is of the same order of magnitude as the molecular sublimation radius,
suggesting that the narrow component of the \FeKa\ line might be 
produced as reflection from the inner walls of the molecular torus.
If so, because of light travelling effects, the intensity of this component 
has to be constant on years time-scales.
We assume the presence of a constant narrow \FeKa\ line 
(as observed by \chandra\ HETG) and add a second, resolved 
(now observed to be broader $\sigma=0.22\pm0.04$ keV), 
\FeKa\ component with EW$=42^{+9}_{-4}$ eV. 
There is excess emission at 7.06 keV, consistent with being 
produced (at least in part) by the associated Fe K$\beta$ emission.

\item{} For the first time reverberation mapping of the resolved 
component of the \FeKa\ line on timescales of several days-years
was successfully performed. 
The intensity of the resolved \FeKa\ component shows a significant 
($\sim4 \sigma$) 1-to-1 correlation with the 3-10 keV flux variability,
however the EW stays constant during the 9 years \xmm\ 
observed the source. 
The robustness of this result is confirmed by the results of the rms
spectra which, in a model independent way, show an excess of 
variability at E$=6.45\pm0.08$ keV. This excess of variability is 
consistent with being the resolved component of the \FeKa\ line 
($\sigma ~ \lsimeq ~ 0.2$ keV) varying in such a way as to 
keep a constant EW$=71\pm36$ eV.  
No measurable lag of the reflected component is observed. 

\item{} The width of the resolved component of the \FeKa\ line 
suggests an origin between 300 and 1000 r$_{\rm g}$ from the 
BH. This location is consistent with the observed \FeKa\ variability 
on days to a week timescale and the lack of measurable 
lag. The lack of a relativistic red wing of the \FeKa\ line suggests
an inner radius for the line production larger than several tens of 
r$_{\rm g}$ ($\sim40$ r$_{\rm g}$). 

\item{} The EW$=42^{+9}_{-4}$ eV of the resolved \FeKa\ line 
suggests a larger covering factor of the 
primary X-ray sources (assumed to have altitudes of few r$_{\rm g}$
above the BH) compared to the one expected from a flat disc annulus,
indicating a possible azimuthal distribution above the disc of the 
reflecting material. A possibility is that the material producing the 
resolved \FeKa\ emission might be in the form of clouds,
perhaps associated to the inner BLR (see Costantini et al. 2012).
This geometry is further reinforced by the consistency between the 
\FeKa\ line width ($\sigma=0.22$ keV) and those from the broadest 
components of the UV broad emission lines (Kriss et al. 2011). 
We also observe that the location of the reverberating \FeKa\ emission 
does not correspond to that of any X-ray or UV absorption components (Detmers et al. 
2011; Kaastra et al. 2012; Kriss et al. 2011; 2012; Ebrero et al. 2011). 

\item{} Significant, but weak (15-20 eV) ionised Fe K emission is 
observed. The ionised emission can be fit equally well with two 
narrow emission lines (from both \Fevc\ and \Fevs), possibly 
from a photo-ionised or collisionally ionised gas, or by a single broad 
relativistic emission line (either \Fevc\ or \Fevs). 
We note that the source high Eddington ratio suggests the presence 
of a standard thin $\alpha$-disc down to a few r$_{\rm g}$ from 
the BH. However, the neutral \FeKa\ line has no redshifted wing with 
no neutral emission closer than $\sim40$ r$_{\rm g}$ from the BH.
This suggests that the surface of the inner accretion disc in Mrk 509 might 
be highly ionised. For these reasons, although the two interpretations 
for the origin of the ionised Fe K emission are equivalent on a statistical 
ground, we slightly prefer the latter interpretation on physical grounds.
The picture of an higher ionised disc in the inner few tens of r$_{\rm g}$ 
from the BH and less ionised outside is in line with the presence of 
a compact hard X-ray corona, providing there a high flux of hard X-ray photons,
and a soft more extended one, as proposed by Petrucci et al. (2012). 

\item{} A highly ionised, medium outflow velocity ($v\sim0.048\pm0.013$ c) 
Fe K absorption component detected in previous observations 
(EW$=-13^{+5.9}_{-2.9}$ eV at $\sim~4~\sigma$ significance) appears 
much weaker (EW$=-3.2^{+2.6}_{-2.8}$ eV, $\sim4$ times weaker), 
if not absent, during the 2009 campaign.

\item{} Previous \xmm\ observations showed evidence for 
highly ionised high outflow velocity ($v\sim0.05-0.2$ c) absorbers
based on a total exposure of $\sim300$ ks.
We find no convincing ($>3 \sigma$) evidence for these features 
during the 2009 \xmm\ long (600 ks) monitoring campaign.

\end{itemize}

\section*{Acknowledgments}

This work is based on observations obtained with \xmm, an ESA science
mission with instruments and contributions directly funded by ESA Member States
and the USA (NASA). We thank the anonymous referee for very helpful 
comments. GP acknowledges support via an EU Marie 
Curie Intra-European Fellowship under contract no. FP7-PEOPLE-2009-IEF-254279.
SRON is supported financially by NWO, the Netherlands Organization for Scientific 
Research. P.-O. Petrucci acknowledges financial support from CNES and the French 
GDR PCHE. M. Cappi, M. Dadina, S. Bianchi, and G. Ponti acknowledge financial support from
contract ASI-INAF n. I/088/06/0. N. Arav and G. Kriss gratefully acknowledge
support from NASA/XMM-Newton Guest Investigator grant NNX09AR01G. Support for
HST Program number 12022 was provided by NASA through grants from the Space
Telescope Science Institute, which is operated by the Association of
Universities for Research in Astronomy, Inc., under NASA contract NAS5-26555. E.
Behar was supported by a grant from the ISF. P. Lubi\'nski has been supported by the
Polish MNiSW grants N N203 581240 and 362/1/N-INTEGRAL/2008/09/0. M. Mehdipour
acknowledges the support of a PhD studentship awarded by the UK Science \&
Technology Facilities Council (STFC). K. Steenbrugge acknowledges the support of Comit\'e
Mixto ESO - Gobierno de Chile.

\newpage
\begin{landscape}
\begin{table} 
\begin{center}
\small
\begin{tabular}{c | c | c c c c | c | c c  c c | c}
\hline                      
\multicolumn{12}{c}{~}\\
\multicolumn{12}{c}{{\bf 2009 campaign - summed spectrum - 1 component of the \FeKa\ line}} \\
\multicolumn{12}{c}{~}\\
\multicolumn{2}{l}{~} & \multicolumn{4}{c}{Fe K$\alpha$} & \multicolumn{1}{c}{Fe K$\beta$} & \multicolumn{2}{c}{Fe {\sc xxv}} & \multicolumn{2}{c}{Fe {\sc xxvi}} \\ 
\hline
Model & $\Gamma$ & E$_{\rm Fe K\alpha}$ & $\sigma_{\rm Fe K\alpha}$ & N$_{\rm Fe K\alpha}$\dag & EW$_{\rm Fe K\alpha}$ & EW$_{\rm Fe K\beta}$ & N$_{\rm \Fevc}$\dag   &\multicolumn{1}{c|}{EW$_{\rm \Fevc}$} & N$_{\rm \Fevs}$\dag & EW$_{\rm \Fevs}$ & $\chi^2$/d.o.f. \\ 
          &                    & (keV)                          & (keV)                                    & (ph cm$^{-2}$ s$^{-1}$)     & (eV)                                & (eV)                              &(ph cm$^{-2}$ s$^{-1}$)&\multicolumn{1}{c|}{(eV)}                       & (ph cm$^{-2}$ s$^{-1}$)&(eV) \\
& & & & & & & & \multicolumn{1}{c|}{~} & & &  \\
(1) & $1.66\pm0.01$&$6.43\pm0.01$          & $0.14\pm0.02$                     & $3.8\pm0.3$                     & $70\pm5$                        &                                      &                                     &\multicolumn{1}{c|}{~}                               &                                  &                                &  1311.7/1194 \\ 
& & & & & & & & \multicolumn{1}{c|}{~} & & &  \\
(2) & $1.66\pm0.01$ & $6.43\pm0.01$          & $0.14\pm0.02$                     & $3.8\pm0.3$                     & $70\pm5$                        & $20\pm3\diamond$       &                                     &\multicolumn{1}{c|}{~}                                &                                  &                                &  1244.8/1193 \\ 
& & & & & & & & \multicolumn{1}{c|}{~} & & &  \\
(3) & $1.66\pm0.01$ & $6.415\pm0.012$       &  $0.092\pm0.012$               & $3.2^{+0.3}_{-0.2}$           & $58.3^{+5.0}_{-2.9}$      & $10.4\ddag$                & $0.54^{+0.14}_{-0.18}$ &\multicolumn{1}{c|}{9.8}                       & $0.29^{+0.11}_{-0.13}$ & 6 &  1219.8/1192 \\ 
\multicolumn{12}{c}{~}\\
\multicolumn{2}{l}{~} & \multicolumn{4}{c}{Fe K$\alpha$} & \multicolumn{1}{c}{Fe K$\beta$} & \multicolumn{4}{c}{Diskline}  \\ 
\hline   
     & $\Gamma$      & E$_{\rm Fe K\alpha}$ & $\sigma_{\rm Fe K\alpha}$ & N$_{\rm Fe K\alpha}$\dag & EW$_{\rm Fe K\alpha}$ & EW$_{\rm Fe K\beta}$ & incl.            & emissiv.       & N$_{\rm diskline}$\dag  & EW$_{\rm diskline}$ & $\chi^2$/d.o.f. \\ 
     &                          & & & & & & (deg.)        &                      & (ph cm$^{-2}$ s$^{-1}$)& (eV)                          & \\
(4) &$1.67\pm0.01$ & $6.415\pm0.011$       &  $0.084\pm0.015$               & $2.7\pm0.2$                  & $50.0\pm3.8$             & $8.2\ddag$                          & $33^{+4}_{-7}$ & $-2.2^{+0.5}_{-0.3}$ & $1.7\pm0.5$      & $36\pm10$       &  1216.0/1191 \\
& & & & & & & & & & &   \\
(5) &$1.67\pm0.01$ & $6.425\pm0.011$       &  $0.094\pm0.018$               & $2.9\pm0.3$                  & $53.8^{+6.2}_{-4.7}$  & $8.6\ddag$                          & $18^{+4}_{-8}$ & $-2.8\pm0.3$ & $1.8^{+0.1}_{-0.5}$      & $40^{+22}_{-11}$ & 1219.4/1191 \\ 
& & & & & & & & & & &   \\
\hline   
\end{tabular}
\small
\caption{Best fit results of the summed EPIC-pn spectrum of the 10 \xmm\ observations performed during the 2009 campaign. 
$\diamond$: assuming $\sigma_{\rm Fe~K\beta}=\sigma_{\rm Fe~K\alpha}$; \ddag: assuming $\sigma_{\rm Fe~K\beta}=\sigma_{\rm Fe~K\alpha}$ and N$_{\rm Fe~K\beta}=0.15\times$ N$_{\rm Fe~K\alpha}$; \dag: in units of $10^{-5}$ ph cm$^{-2}$ s$^{-1}$.}
\begin{itemize}
\item Model (1) {\it Single FeK$\alpha$}: Power law~+~Gaus$_{\rm Fe K\alpha}$ 
\item Model (2) {\it Single FeK$\alpha+\beta$}: Power law~+~Gaus$_{\rm Fe K\alpha}$~+~Gaus$_{\rm Fe K\beta}$ 
\item Model (3) {\it Photo-ionised gas + Single FeK$\alpha+\beta$}: Power law~+~Gaus$_{\rm Fe K\alpha}$~+~Gaus$_{\rm Fe K\beta}$~+~Gaus$_{\rm Fe XXV}$~+~Gaus$_{\rm Fe XXVI}$  
\item Model (4) {\it Broad line$_{\rm Fe XXV}$ + Single FeK$\alpha+\beta$}: Power law~+~Gaus$_{\rm Fe K\alpha}$~+~Gaus$_{\rm Fe K\beta}$~+~Diskline$_{\rm Fe XXV}$
\item Model (5) {\it Broad line$_{\rm Fe XXVI}$ + Single FeK$\alpha+\beta$}: Power law~+~Gaus$_{\rm Fe K\alpha}$~+~Gaus$_{\rm Fe K\beta}$~+~Diskline$_{\rm Fe XXVI}$
\end{itemize}
\label{tabMean}
\end{center}                   
\end{table}
\clearpage
\begin{table} 
\hspace{-2.5cm}
\begin{center}
\hspace{-2.5cm}
\tiny
\begin{tabular}{c | c | c c c | c c c | c c | c c | c c | c}
\hline                      
\multicolumn{15}{c}{~} \\
\multicolumn{15}{c}{{\bf 2009 campaign - summed spectrum - 2 components of the \FeKa\ line}} \\
\multicolumn{15}{c}{~} \\
\multicolumn{2}{c}{~} & \multicolumn{3}{c}{Narrow Fe K$\alpha$} & \multicolumn{3}{c}{Resolved Fe K$\alpha$} & \multicolumn{2}{c}{Fe {\sc xxv}} & \multicolumn{2}{c}{Fe {\sc xxvi}} & \multicolumn{2}{c}{Absorption line} \\ 
\hline
Model & $\Gamma$       & E$_{\rm Fe K\alpha}$ & N$_{\rm Fe K\alpha N}$\dag & EW$_{\rm Fe K\alpha N}$& $\sigma_{\rm Fe K\alpha R}$ & N$_{\rm Fe K\alpha R}$\dag & EW$_{\rm Fe K\alpha R}$ & N$_{\rm \Fevc}$\dag    & EW$_{\rm \Fevc}$        & N$_{\rm \Fevs}$\dag & EW$_{\rm \Fevs}$ & E$_{\rm abs}$ & N$_{\rm abs}\dag$ & $\chi^2$/d.o.f. \\ 
           &                         & (keV)                          & (phcm$^{-2}$ s$^{-1}$)         & (eV)                                & (keV)                                        &(ph cm$^{-2}$ s$^{-1}$)        & (eV)                                  & (ph cm$^{-2}$ s$^{-1}$) & (eV)                              & (ph cm$^{-2}$ s$^{-1}$)& (eV)                       &                         &  (ph cm$^{-2}$ s$^{-1}$~:~eV)& \\
(6)      & $1.67\pm0.01$& $6.420\pm0.010$       &$1.5$\dag                              & $27$\dag                           & $0.22\pm0.04$                       & $2.33^{+0.38}_{-0.27}$           & $42^{+9}_{-4}$                    & $0.21\pm0.15$              & $3.9\pm2.7$                & $0.20\pm0.13$          & $4.2\pm2.6$                  & $7.31\pm0.1$  & $-0.15\pm0.12 : 3.4\pm2.7$ & 1197.9/1190 \\ 
& & & & & & & & & & & & & &  \\
\hline   
\multicolumn{15}{c}{~} \\
\multicolumn{15}{c}{{\bf 2000 - 2006 \xmm\ observations - summed spectrum - 2 components of the \FeKa\ line}} \\
\multicolumn{15}{c}{~} \\
\hline   
(6)      & $1.67\pm0.01$& $6.427\pm0.014$       &$1.5$\dag                              & $33$\dag                           & $0.21^{+0.12}_{-0.07}$           & $1.50^{+0.60}_{-0.38}$        & $33^{+6}_{-7}$                  & $<0.47$                          & $<10.4$                        & $0.37\pm0.18$          &$9.5\pm4.5$ & $7.34\pm0.05$ & $-3.6\pm0.2 : 10.1\pm3.8$& 1212.3/1190 \\
& & & & & & & & & & & & & &  \\
\hline   
\end{tabular}
\caption{\dag: in units of $10^{-5}$ ph cm$^{-2}$ s$^{-1}$.}
\label{2lines1}
\end{center}                   
\end{table}
\begin{itemize}
\item Model (6) {\it Photo-ionised gas + Double FeK$\alpha$ + Absorption line}: Power law~+~Gaus$_{\rm Fe K\alpha N}$~+~Gaus$_{\rm Fe K\alpha R}$~+~Gaus$_{\rm Fe K\beta N}$~+~Gaus$_{\rm Fe K\beta R}$~+~Gaus$_{\rm Fe XXV}$~+~Gaus$_{\rm Fe XXVI}$~-~AbsGaus
\end{itemize}

\begin{table} 
\begin{center}
\begin{tabular}{c | c | c c c | c c c | c c c | c}
\hline                      
\multicolumn{12}{c}{~} \\
\multicolumn{12}{c}{{\bf 2009 campaign - summed spectrum - 2 components of the \FeKa\ line}} \\
\multicolumn{12}{c}{~} \\
\multicolumn{2}{c}{~} & \multicolumn{3}{c}{Narrow Fe K$\alpha$} & \multicolumn{3}{c}{Resolved Fe K$\alpha$} & \multicolumn{3}{c}{Diskline} &  \\ 
\hline
Model & $\Gamma$       & E$_{\rm Fe K\alpha}$ & N$_{\rm Fe K\alpha N}$\dag & EW$_{\rm Fe K\alpha N}$& $\sigma_{\rm Fe K\alpha R}$ & N$_{\rm Fe K\alpha R}$\dag & EW$_{\rm Fe K\alpha R}$ & $\sigma_{\rm ion.}$       & N$_{\rm ion.}$             & EW$_{\rm ion.}$        &$\chi^2$/d.o.f. \\ 
           &                         & (keV)                          & (ph cm$^{-2}$ s$^{-1}$)           & (eV)                                & (keV)                                        &(ph cm$^{-2}$ s$^{-1}$)          & (eV)                                  & (keV)                             & (ph cm$^{-2}$ s$^{-1}$) & (eV)                          &    \\
(7)     & $1.67\pm0.01$& $6.416\pm0.011$       &$1.5$\dag                               & $27$\dag                          & $0.23^{+0.05}_{-0.04}$           & $2.02^{0.54}_{-0.32}$           & $36^{+9}_{-6}$                 &$0.23^{+0.1}_{-0.06}$    &$0.79^{+0.66}_{-0.40}$& $15^{+12}_{-8}$        &1203.5/1192 \\ 
         &                          &                                    &                                               &                                          &                                                 &                                               &                                          & r$_{\rm in}$                   & N$_{\rm disk}$             & EW$_{\rm ion.}$        & $\chi^2$/d.o.f. \\ 
         &                          &                                    &                                               &                                          &                                                 &                                               &                                          & (r$_{\rm g}$)                 & (ph cm$^{-2}$ s$^{-1}$) & (eV)                         & \\          
(8)     & $1.67\pm0.01$& $6.425\pm0.013$       &$1.5$\dag                               & $27$\dag                          & $0.21^{+0.07}_{-0.03}$           & $1.92^{+0.85}_{-0.27}$          & $34^{+15}_{-5}$            &$27^{+41}_{-20}$           &$0.92^{+0.46}_{-0.36}$& $20^{+6}_{-9}$           &1204.7/1192 \\ 
& & & & & & & & & & & \\
\hline   
\end{tabular}
\caption{\dag: in units of $10^{-5}$ ph cm$^{-2}$ s$^{-1}$.}
\label{2lines2}
\end{center}                   
\end{table}
\begin{itemize}
\item Model (7) {\it Broad ionised Gaussian$_{\rm Fe XXVI}$ + Double FeK$\alpha$}: Power law~+~Gaus$_{\rm Fe K\alpha N}$~+~Gaus$_{\rm Fe K\alpha R}$~+~Gaus$_{\rm Fe K\beta N}$~+~Gaus$_{\rm Fe K\beta R}$~+~Gaus$_{\rm Fe XXV}$~+~Gaus$_{\rm Fe XXVI}$~-~AbsGaus
\item Model (8) {\it Ionised-disc-line$_{\rm Fe XXVI}$ + Double FeK$\alpha$ }: Power law~+~Gaus$_{\rm Fe K\alpha N}$~+~Gaus$_{\rm Fe K\alpha R}$~+~Gaus$_{\rm Fe K\beta N}$~+~Gaus$_{\rm Fe K\beta R}$~+~Gaus$_{\rm Fe XXV}$~+~Gaus$_{\rm Fe XXVI}$~-~AbsGaus
\end{itemize}

\end{landscape}

\end{document}